\documentclass[psfig]{aa}
\input psfig.tex

\begin{document}                                                                

\twocolumn


   \title{Observational Constraints on General Relativistic Energy
	 Conditions, Cosmic Matter Density and Dark Energy from X-Ray
	 Clusters of Galaxies and Type-Ia Supernovae}

   \titlerunning{Observational Constraints on General Relativistic
Energy Conditions}

   \author{P. Schuecker$^{(1)}$, R. R. Caldwell$^{(2)}$,
H. B\"ohringer$^{(1)}$, C.A. Collins$^{(3)}$, L. Guzzo$^{(4)}$,
N. N. Weinberg$^{(5)}$}

   \authorrunning{Schuecker et al.}

   \offprints{Peter Schuecker\\ peters@mpe.mpg.de}

   \institute{             
    $^{(1)}$ Max-Planck-Institut f\"ur extraterrestrische Physik,
             Giessenbachstra{\ss}e 1, 85740 Garching, Germany\\ 
    $^{(2)}$ Department of Physics \& Astronomy, Dartmouth College,
 Hanover, NH\,03755, USA\\
    $^{(3)}$ Astrophysics Research Institute, Liverpool John Moores
 University, Twelve Quays House, Egerton Wharf, Birkenhead CH41 1LD,
 Great Britain\\
    $^{(4)}$ INAF-Osservatorio di Brera, via Bianchi, 22055 Merate (LC), Italy\\
    $^{(5)}$ California Institute of Technology, Mail Code 130-33,
Pasadena, CA 91125, USA
}

   \date{Received  ; accepted }                         	
   
   \markboth{Observational Constraints on General Relativistic Energy
Conditions}{}

\abstract{New observational constraints on the cosmic matter density
$\Omega_{\rm m}$ and an effectively redshift-independent equation of state
parameter $w_{\rm x}$ of the dark energy are obtained while simultaneously
testing the strong and null energy conditions of general relativity on
macroscopic scales. The combination of REFLEX X-ray cluster and
type-Ia supernova data shows that for a flat Universe the strong
energy condition might presently be violated whereas the null energy
condition seems to be fulfilled.  This provides another observational
argument for the present accelerated cosmic expansion and the absence
of exotic physical phenomena related to a broken null energy
condition. The marginalization of the likelihood distributions is
performed in a manner to include a large fraction of the recently
discussed possible systematic errors involved in the application of
X-ray clusters as cosmological probes. This yields for a flat
Universe, $\Omega_{\rm m}=0.29^{+0.08}_{-0.12}$ and
$w_{\rm x}=-0.95^{+0.30}_{-0.35}$ ($1\sigma$ errors without cosmic
variance). The scatter in the different analyses indicates a quite
robust result around $w_{\rm x}=-1$, leaving little room for the
introduction of new energy components described by quintessence-like
models or phantom energy. The most natural interpretation of the data
is a positive cosmological constant with $w_{\rm x}=-1$ or something like
it.  }

\maketitle

\section{Introduction}\label{INTRO}

Measurements of the cosmic microwave background (CMB) temperature
anisotropies (e.g. Stompor et al. 2001, Netterfield et al. 2002, Pryke
et al. 2002, Scott 2002, Sievers et al. 2002, Spergel et al. 2003),
the redshift-distance relation of type-Ia supernovae (Riess et
al. 1998, Perlmutter et al. 1999), the counts of galaxy clusters
(e.g. Bahcall \& Fan 1998, Borgani et al. 2001, Reiprich \&
B\"ohringer 2002), etc., suggest that we live in a dark-energy
dominated Universe during a phase of accelerated cosmic expansion.
Perhaps the simplest resolution is to resort to Einstein's
cosmological constant $\Lambda$. Postulating a constant leaves many
questions unanswered, however, relating to the nature of the particle
physics vacuum and the (approximate) coincidence in the energy density
of dark energy and dark matter today. For this finely-tuned constant
the answer would seem to lie in the initial conditions. An alternative
hypothesis is to consider a time-evolving dark energy, while assuming
that any cosmological term is either zero or negligible. For a
time-evolving inhomogeneous field (see e.g. Ratra \& Peebles 1988,
Wetterich 1988, Caldwell et al. 1998, Caldwell 2002) the aim is to
understand the coincidence in terms of dynamics.

A central r\^{o}le in these studies is assumed by the phenomenological
ratio $w=p/\rho c^2$ between the pressure $p$ of the unknown energy
component and its rest energy density $\rho$. In most investigations
the parameter space of $w$ is restricted to $w\ge -1$ (exceptions are
Caldwell 2002, Hannestad \& M\"ortsell 2002, and Melchiorri et
al. 2002) by assuming that the so-called null energy condition of
general relativity should be fulfilled on macroscopic scales. However,
for energy conditions no strict mathematical proofs exist hitherto,
and their validity is not more than a conjecture. Therefore, the
present investigation takes one step back, using $w$ itself to test
the energy condition and finding the $w$ value which represents the
observational data best.
 
Before we describe the test and its application to astronomical data
we briefly review the significance of the energy conditions in
cosmology relevant for the present work and their relation to $w$
(Sect.\,\ref{ENERGY}). The test is outlined in Sect.\,\ref{TEST}. In
Sect.\,\ref{OBS} the observational material used for the test is
described. The results are presented in Sect.\,\ref{RESULTS} and
discussed in Sect.\,\ref{DISCUSS}. In the following we define the
Hubble constant in units of $h=H_0/(100\,{\rm km}\,{\rm s}^{-1}\,{\rm
Mpc}^{-1})$.

\section{Energy conditions and cosmology}\label{ENERGY}

\begin{figure}
\vspace{-1.0cm}
\centerline{\hspace{-0.5cm}
\psfig{figure=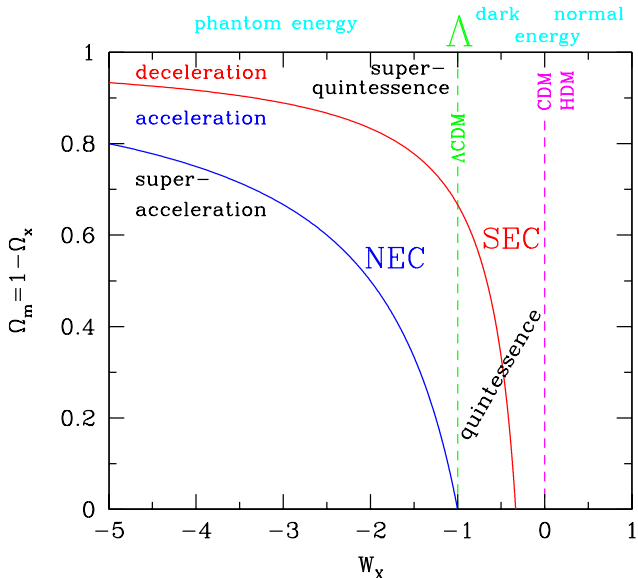,height=9.5cm,width=9.5cm}}
\vspace{-1.0cm}
\caption{\small The null energy condition (NEC) and the strong energy
condition (SEC) for a flat FRW spacetime at redshift $z=0$ with
negligible contributions from relativistic particles in the parameter
space of the normalized cosmic density $\Omega_{\rm m}$ of baryonic and
non-baryonic matter and the equation of state parameter $w_{\rm x}$ of a
presently unknown energy component. The NEC and SEC curves are
computed with Eqs.\,(\ref{NEC}) and (\ref{SEC}), respectively. The
sector between the two curves gives accelerated growing scale factors
where NEC is fulfilled but SEC is violated. In the sector below the
NEC curve all energy conditions of general relativity are violated and
the scale factor shows a super-accelerated increase. In the sector
above SEC all energy conditions are fulfilled (especially also SEC)
and no accelerated cosmic expansion is expected. The vertical line at
$w_{\rm x}=-1$ mark models with Einstein's cosmological constant
($\Lambda>0$, especially Cold Dark Matter models $\Lambda$CDM) and
devides the parameter space into the quintessence-like sector
($-1<w_{\rm x}<0$) based on dark energy and the super-quintessence sector
($w_{\rm x}<-1$) based on phantom energy. For $0 \le w_{\rm x}\le 1$ ordinary
energy might be expected in the form of Cold Dark Matter (CDM), Hot
Dark Matter (HDM) etc.}
\label{FIG_SCATCH}
\end{figure}

Assumptions on energy conditions form the basis for the well-known
singularity theorems (Hawking \& Ellis 1973), censorship theorems
(e.g. Friedman et al. 1993) and no-hair theorems (e.g. Mayo \&
Bekenstein 1996). Quantized fields violate all local point-wise energy
conditions (Epstein et al. 1965). In the present investigation we are,
however, concerned with observational studies on macroscopic scales
relevant for cosmology where $\rho$ and $p$ are expected to behave
classically. Normal matter in the form of baryons and non-baryons, or
relativistic particles like photons and neutrinos satisfy all standard
energy conditions. The two energy conditions discussed below are given
in a simplified form (for more details see Wald 1984 and Barcel\'{o}
\& Visser 2000).

The {\it strong energy condition} (SEC): $\rho+3p/c^2\ge0$ {\it and}
$\rho+p/c^2\ge0$, derived from the more general condition
$R_{\mu\nu}v^\mu v^\nu\ge0$, where $R_{\mu\nu}$ is the Ricci tensor
for the geometry and $v^\mu$ a timelike vector. The simplified
condition is valid for diagonalizable energy-momentum tensors which
describe all observed fields with non-zero rest mass and all zero rest
mass fields except some special cases (see Hawking \& Ellis 1973). The
SEC ensures that gravity is always attractive. Certain singularity
theorems (e.g., Hawking \& Penrose 1970) relevant for proving the
existence of an initial singularity in the Universe need an attracting
gravitational force and thus assume SEC. Violations of this condition
as discussed in Visser (1997) allows phenomena like inflationary
processes expected to take place in the very early Universe or a
moderate late-time accelerated cosmic expansion as suggested by the
combination of recent astronomical observations (see
Sect.\,\ref{INTRO}). Likewise, phenomena related to $\Lambda>0$ and an
effective version of $\Lambda$ whose energy and spatial distribution
evolve with time ({\it quintessence}: Ratra \& Peebles 1988, Wetterich
1988, Caldwell et al. 1998 etc.)  are allowed consequences of the
breaking of SEC -- but not a prediction.  However, a failure of SEC
seems to have no severe consequences because the theoretical
description of the relevant physical processes can still be provided
in a canonical manner. Phenomenologically, violation of SEC means
$w<-1/3$ for a {\it single} energy component with density $\rho>
0$. For $w\ge -1/3$, SEC is not violated and we have a decelerated
cosmic expansion.

The {\it null energy condition} (NEC): $\rho+p/c^2\ge0$, derived from
the more general condition $G_{\mu\nu}k^\mu k^\nu\ge0$, where
$G_{\mu\nu}$ is the geometry-dependent Einstein tensor and $k^\mu$ a
null vector (energy-momentum tensors as for SEC). Violations of this
condition are recently studied theoretically in the context of
macroscopic traversable wormholes (see averaged NEC: Flanagan \& Wald
1996, Barcel\'{o} \& Visser 2000) and the holographic principle (see
covariant entropy bound, McInnes 2002). The breaking of this criterion
in a finite local region would have subtle consequences like the
possibility for the creation of ``time machines'' (e.g. Morris, Thorne
\& Yurtsever 1988). Violating the energy condition in the cosmological
case is not as dangerous (no threat to causality, no need to involve
chronology protection, etc.), since one cannot isolate a chunk of the
energy to power such exotic objects. Nevertheless, violation of NEC on
cosmological scales could excite phenomena like super-acceleration of
the cosmic scale factor (Caldwell 2002). Theoretically, violation of
NEC would have profound consequences not only for cosmology because
all point-wise energy conditions would be broken. It cannot be
achieved with a canonical Lagrangian {\it and} Einstein
gravity. Phenomenologically, violation of NEC means $w<-1$ for a {\it
single} energy component with $\rho> 0$. The sort of energy related to
this state of a Friedmann-Robertson-Walker (FRW) spacetime is dubbed
{\it phantom energy} and is described by {\it super-quintessence}
models (Caldwell 2002, see also Chiba et al. Yamaguchi 2000). For
$w\ge-1$ NEC is not violated, and the sort of energy is termed {\it
dark energy} and is described by {\it quintessence} or
super-quintessence models.

\section{A test of energy conditions on large scales}\label{TEST}

\begin{figure}
\vspace{-1.5cm}
\centerline{\hspace{-0.25cm}
\psfig{figure=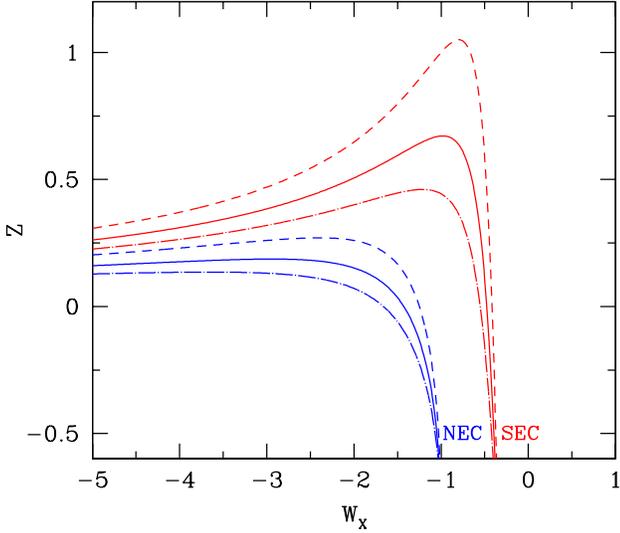,height=9.5cm,width=9.5cm}}
\vspace{-1.0cm}
\caption{\small The null energy condition (NEC) and the strong energy
condition (SEC) for a flat FRW spacetime as a function of redshift $z$
and $w_{\rm x}$ with negligible contributions from relativistic particles
computed with Eq.\,(\ref{EC}). Short dashed curves are computed with
$\Omega_{\rm m}=0.2$, continuous curves with $\Omega_{\rm m}=0.3$, and
dashed-dotted curves with $\Omega_{\rm m}=0.4$. Above the curves the
respective energy conditions are fulfilled.}
\label{FIG_NSEC_Z}
\end{figure}

The lack of verification of the energy conditions on macroscopic
scales suggest that we should first test observationally the degree to
which we can presently trust NEC and SEC. It will be seen that the
test automatically yields those values of the present cosmic matter
density and the equation-of-state parameter of the dark energy which
describe the observational data best.

Our starting point is a FRW spacetime filled with a positive energy
density of unknown nature which can be described as a perfect fluid
($p_{\rm x}\not=0$, $\rho_{\rm x}>0$, $w_{\rm x}$), with contributions from a
pressureless non-baryonic and baryonic fluid ($p_{\rm m}=0$, $\rho_{\rm m}>0$,
$w_m=0$), and from relativistic particles like photons or neutrinos
($p_{\rm r}>0$, $\rho_{\rm r}>0$, $w_r=1/3$).

The different fluids `know about each other' through their common
gravitational effects and through possible explicit couplings of one
fluid to the others (a phenomenological treatment of a local energy
transfer between two cosmic fluids can be found in Gromov et
al. 2002). In the late Universe, a coupling between ordinary matter
and the unknown energy component might remain (e.g. Amendola 2000),
but the resulting effects are difficult to distinguish from
predictions of general relativity plus a cosmological constant (Torres
2002). We thus regard the three fluids mentioned above as effectively
independent substances over the redshift range covered by the
astronomical objects used in our tests. Consequently, for each of the
cosmic substances a {\it local} energy balance holds (e.g. Rindler
2001),
\begin{equation}\label{EE}
\frac{\dot{\rho}}{\rho}\,+\,3\,(\,1\,+\,w\,)\,H\,=\,0\,,
\end{equation}
derived from the twice-contracted Bianchi identity and Einstein's
field equations for a perfect fluid. In (\ref{EE}), $H$ is the Hubble
parameter and a dot denotes a derivative with respect to the cosmic
time. 

We are aware that these pre-assumptions are already quite specific
compared to the usually targeted generality. However, our main focus
is to learn more about the unknown energy component `x' and the cosmic
phenomena related to it. In this sense we proceed further and add all
the energy and pressure sources to get the net equation for the
pressumed multi-component effective cosmic fluid. In this case, the
NEC with its constraint on the passive gravitational mass density
($\rho+p/c^2\ge 0$) reads
\begin{equation}\label{EC1}
\rho_{\rm x}\,+\,\frac{p_{\rm x}}{c^2}\,+\,\rho_{\rm m}\,+\,\rho_{\rm r}\,+\,\frac{p_{\rm r}}{c^2}\,\ge\,0\,,
\end{equation}
or with the equation of state parameter $w_{\rm x}$ of the unknown fluid
defined by $p_{\rm x}\,=\,w_{\rm x}\rho_{\rm x} c^2$, and $p_{\rm r}\,=\,\frac{1}{3}\rho_{\rm r}
c^2$,
\begin{equation}\label{EC2}
\rho_{\rm x}\,+\,w_{\rm x}\rho_{\rm x}\,+\,\rho_{\rm m}\,+\,\rho_{\rm
r}\,+\,\frac{1}{3}\rho_{\rm r}\,\ge\,0\,.
\end{equation}
Using the normalized energy density of matter ($\Omega_{\rm m}$), of
relativistic particles ($\Omega_{\rm r}$) and of the presently unknown
energy component ($\Omega_{\rm x}$), Eq.\,(\ref{EC2}) can be recast into the
inequality
\begin{equation}\label{NEC0}
w_{\rm x}\,\ge\,-\,\frac{\Omega_{\rm m}\,+\,\frac
{4}{3}\Omega_{\rm r}\,+\,\Omega_{\rm x}}{\Omega_{\rm x}}\,,
\end{equation}
NEC for $\Omega_{\rm m},\,\Omega_{\rm r}\ge 0,\,\,\Omega_{\rm x}>0$. Similarily, the SEC
with its additional constraint on the active gravitational mass
density $(\rho+3p/c^2\ge 0)$ corresponds to
\begin{equation}\label{SEC0}
w_{\rm x}\,\ge\,-\,\frac{\Omega_{\rm m}\,+\,2\,\Omega_{\rm r}\,+\,\Omega_{\rm x}}{3\,\Omega_{\rm x}}\,,
\end{equation}
SEC for $\Omega_{\rm m},\,\Omega_{\rm r}\ge 0\,\,\Omega_{\rm x}>0$. Note that the
inclusion of an unknown energy component in the form of a perfect
fluid naturally extends earlier definitions of the SEC where cases
with $\Lambda$-like energies were explicitly excluded. However, the
present definition still follows the basic idea of the SEC because its
validity guarantees that the active gravitational mass of a
multi-component cosmic fluid always leads to an attractive
gravitational effect,
\begin{equation}\label{ACC}
\frac{\ddot{a}}{a}\,=\,-\,\frac{4\pi G}{3}\,\sum_{i=x,m,r}\left(\rho_i\,+\,
\frac{3p_i}{c^2}\right)\,\le\,0\,,
\end{equation}
where $a=1/(1+z)$ is the scale factor of the spacetime, $z$ the
cosmological redshift, and where $\Lambda c^2/3$ is replaced in the
Friedmann-Lema\^{\i}\-tre equation by the general term $8\pi
G\rho_{\rm x}/3$.

For the more restrictive case of a spatially flat FRW geometry and a
negligible contribution from relativistic particles (current estimates
range from $\Omega_{\rm r}=0.001$ to maximal $0.05$, see Turner 2002) we
have
\begin{equation}\label{NEC}
w_{\rm x}\, \ge\, -\,\frac{1}{1\,-\,\Omega_{\rm m}}\,,
\end{equation}
NEC for $\Omega_{\rm m}\ge 0$ and $\Omega_{\rm m}+\Omega_{\rm x}=1$, and
\begin{equation}\label{SEC}
w_{\rm x}\, \ge\, -\,\frac{1}{3(1\,-\,\Omega_{\rm m})}\,,
\end{equation}
SEC for $\Omega_{\rm m}\ge 0$ and $\Omega_{\rm m}+\Omega_{\rm x}=1$. Only for the
unrealistic limit $\Omega_{\rm m}\rightarrow 0$ the NEC restriction
(\ref{NEC}) converges to the frequently adopted threshold $w_{\rm x}({\rm
min})=-1$. For $\Omega_{\rm x}\rightarrow 0$ and thus $\Omega_{\rm m}\rightarrow
1$ one gets $w_{\rm x}({\rm min})\rightarrow-\infty$ and both NEC and SEC
are always fulfilled. The link between $w_{\rm x}$ and observable quantities
like distances, volumes etc. is given by equations of the form 
\begin{eqnarray}\label{HUBBLE}
\left[\frac{H(z)}{H_0}\right]^2
\,=\,\Omega_{\rm x}\,f(z,w_{\rm x})\,+\,\Omega_{\rm m}\,(1+z)^3\,+\nonumber\\
\quad\Omega_{\rm r}\,(1+z)^4\,+\,
(\,1\,-\,\Omega_{\rm x}\,-\,\Omega_{\rm m}\,-\,\Omega_{\rm r}\,)\,(\,1\,+\,z\,)^2\,,
\end{eqnarray}
which relate $H(z)$ to $w_{\rm x}$, $\Omega_{\rm m}$, $\Omega_{\rm r}$ and
$\Omega_{\rm x}$. The redshift-dependency of the latter energy component in
(\ref{HUBBLE}) can be obtained from the integration of (\ref{EE}) and
is
\begin{equation}\label{FWZ}
f(z,w_{\rm x})\,=\,\exp\left\{3\,\int_0^z\,[1+w_{\rm x}(z')]\,d\ln(1+z')\right\}\,.
\end{equation}
For simplicity we concentrate on a redshift-independent constant
$w_{\rm x}$, so that (\ref{FWZ}) leads to $f(z,w_{\rm x})=(1+z)^{3(1+w_{\rm x})}$. For
$w_{\rm x}=-1$ we have $f=1$ and $\Omega_{\rm x}$ corresponds to the normalized
cosmological constant $\Omega_\Lambda$ (Fig.\,\ref{FIG_SCATCH}).

For constant $w_{\rm x}$, the redshift-dependency of (\ref{NEC}, \ref{SEC})
can be found by replacing the present matter density $\Omega_{\rm m}$ by
$\Omega_{\rm m}(z)=\Omega_{\rm m}/[\Omega_{\rm m}+(1-\Omega_{\rm m})(1+z)^{3w_{\rm x}}]$ leading to
the condition
\begin{equation}\label{EC}
1\,+\,z\,\ge\,\left[\,-\left(\,1\,+\,\frac{w_{\rm x}}{\gamma}\,\right)\,\frac{1-\Omega_{\rm m}}{\Omega_{\rm m}}\,\right]^{-\frac{1}{3w_{\rm x}}}\,,
\end{equation}
with $\gamma=1$ and $\frac{1}{3}$ for NEC and SEC, respectively, and
$0<\Omega_{\rm m}<1$ and $w_{\rm x}<-\gamma$. Above the redshift limit given in
Eq.\,(\ref{EC}) the respective energy condition is
fulfilled. Fig.\,\ref{FIG_NSEC_Z} suggests that for $z\gg 1$
(excluding cosmic epochs with e.g. dominating scalar fields like in a
possible inflationary phase) and for reasonable present $\Omega_{\rm m}$
values, NEC and SEC are always fulfilled -- independent of the value
of $w_{\rm x}$. It is also seen that the frequently adopted lower
NEC-threshold of $w_{\rm x}({\rm min})=-1$ is recovered at $z\ll 0$, that
is, in the distant future. Note, however, that at extreme redshifts
the constancy of $w_{\rm x}$ is expected to be a poor approximation,
suggesting a failure of our considerations in these extreme
$z$-ranges.

\begin{figure*}
\vspace{-0.0cm}
\centerline{\hspace{-10.5cm}
\psfig{figure=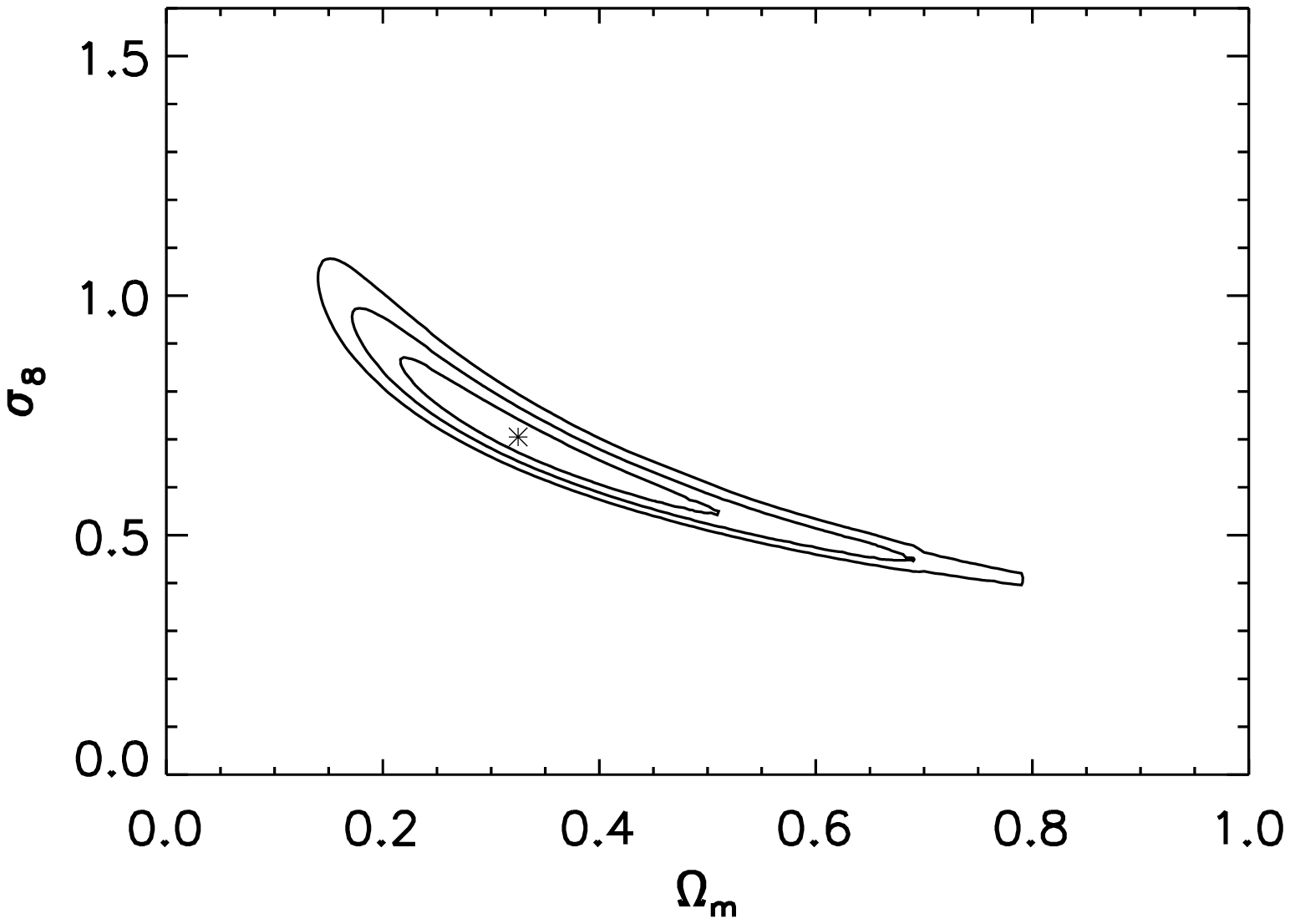,height=5.5cm,width=8.5cm}}
\vspace{-5.5cm}
\centerline{\hspace{ 8.2cm}
\psfig{figure=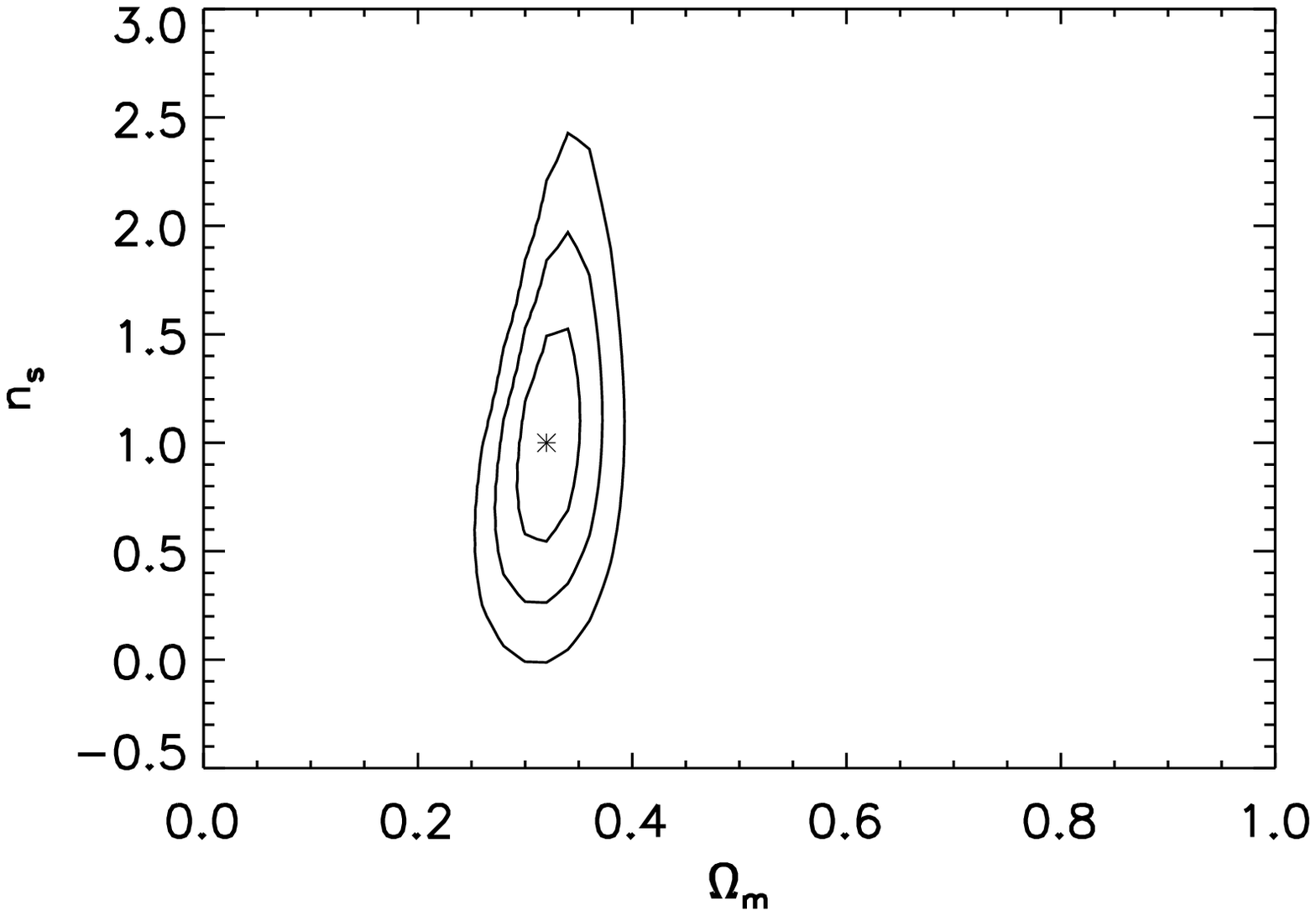,height=5.5cm,width=8.5cm}}
\vspace{-0.0cm}
\centerline{\hspace{-10.5cm}
\psfig{figure=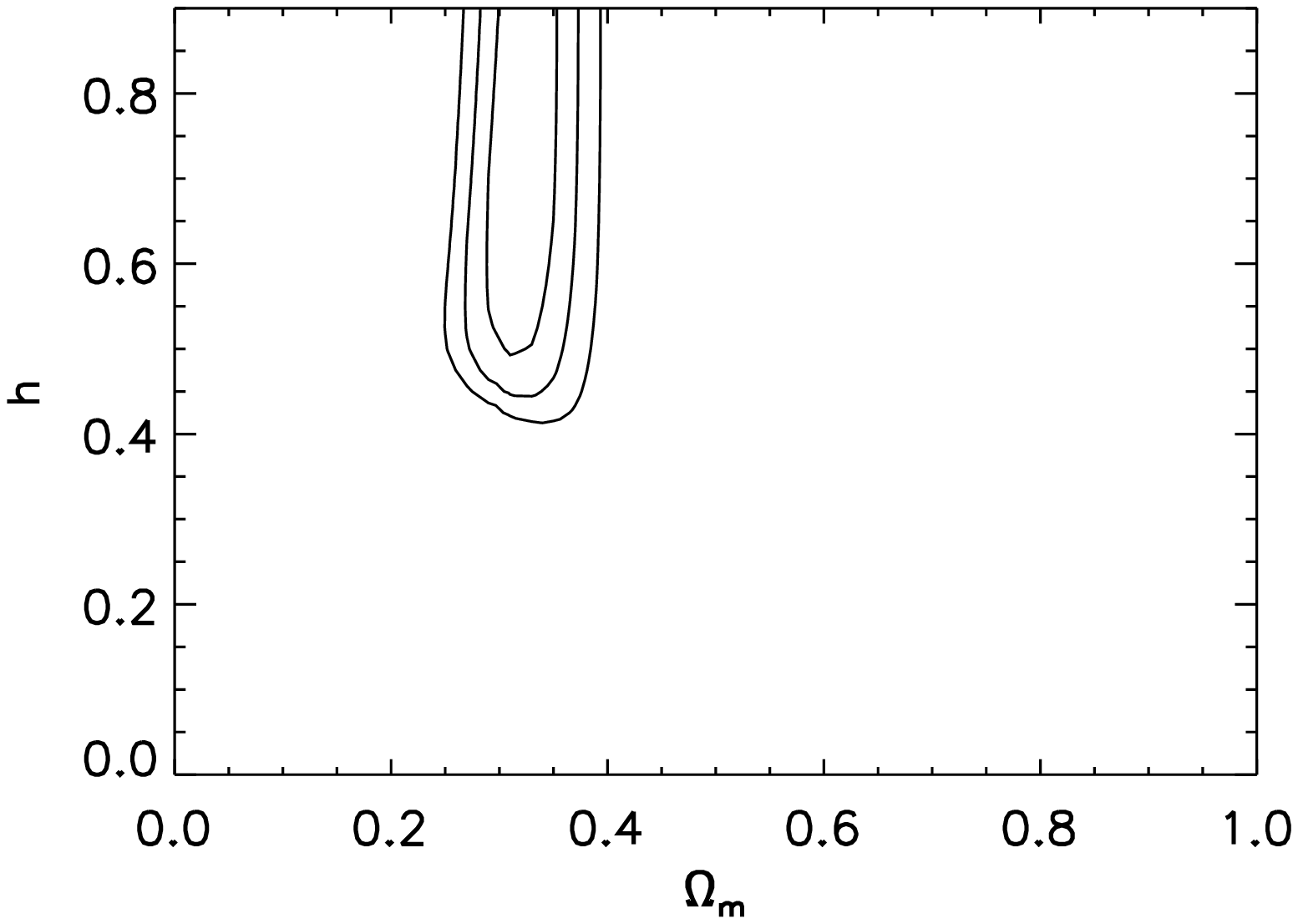,height=5.5cm,width=8.5cm}}
\vspace{-5.5cm}
\centerline{\hspace{ 8.2cm}
\psfig{figure=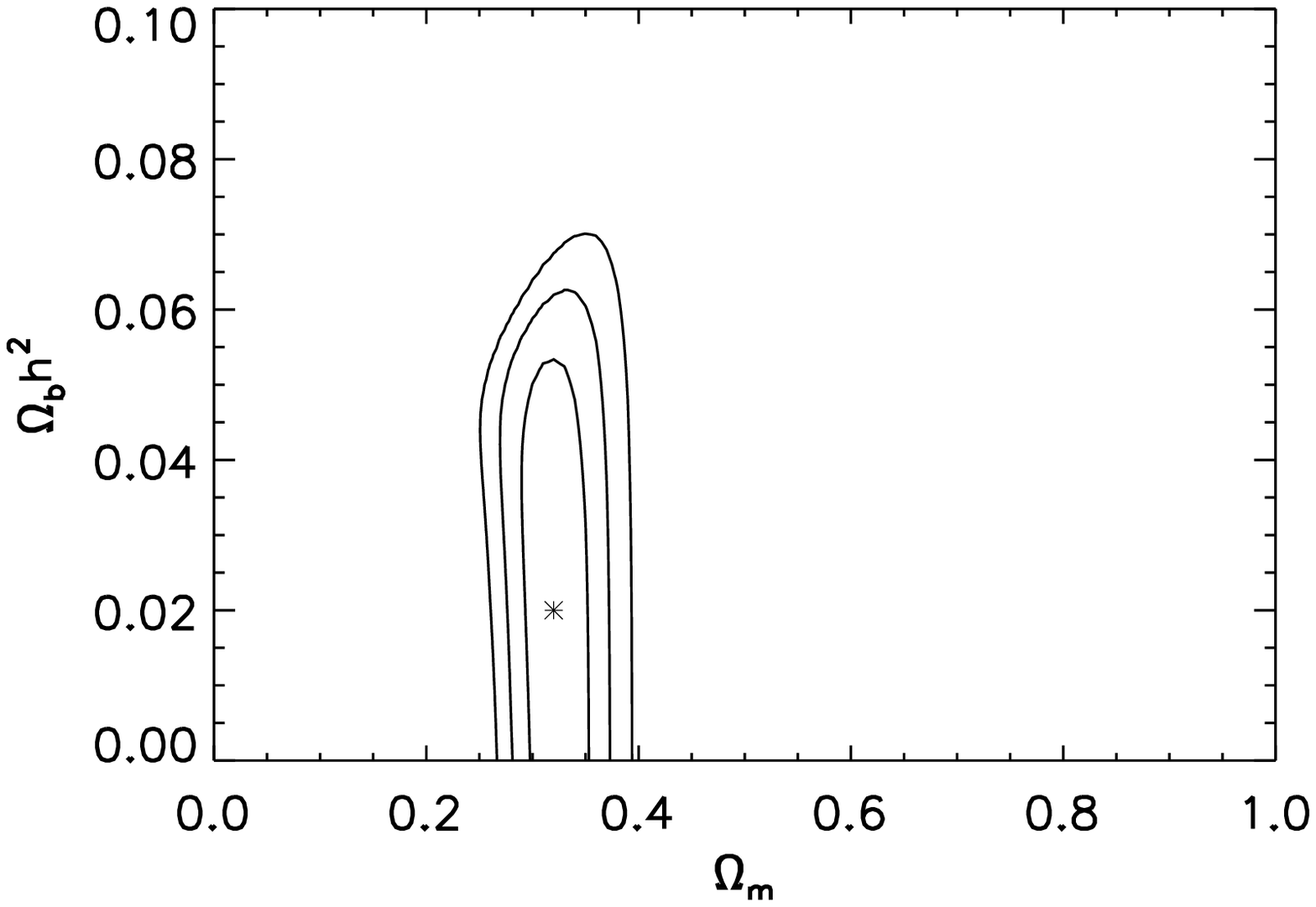,height=5.5cm,width=8.5cm}}
\vspace{-0.0cm}
\centerline{\hspace{-10.5cm}
\psfig{figure=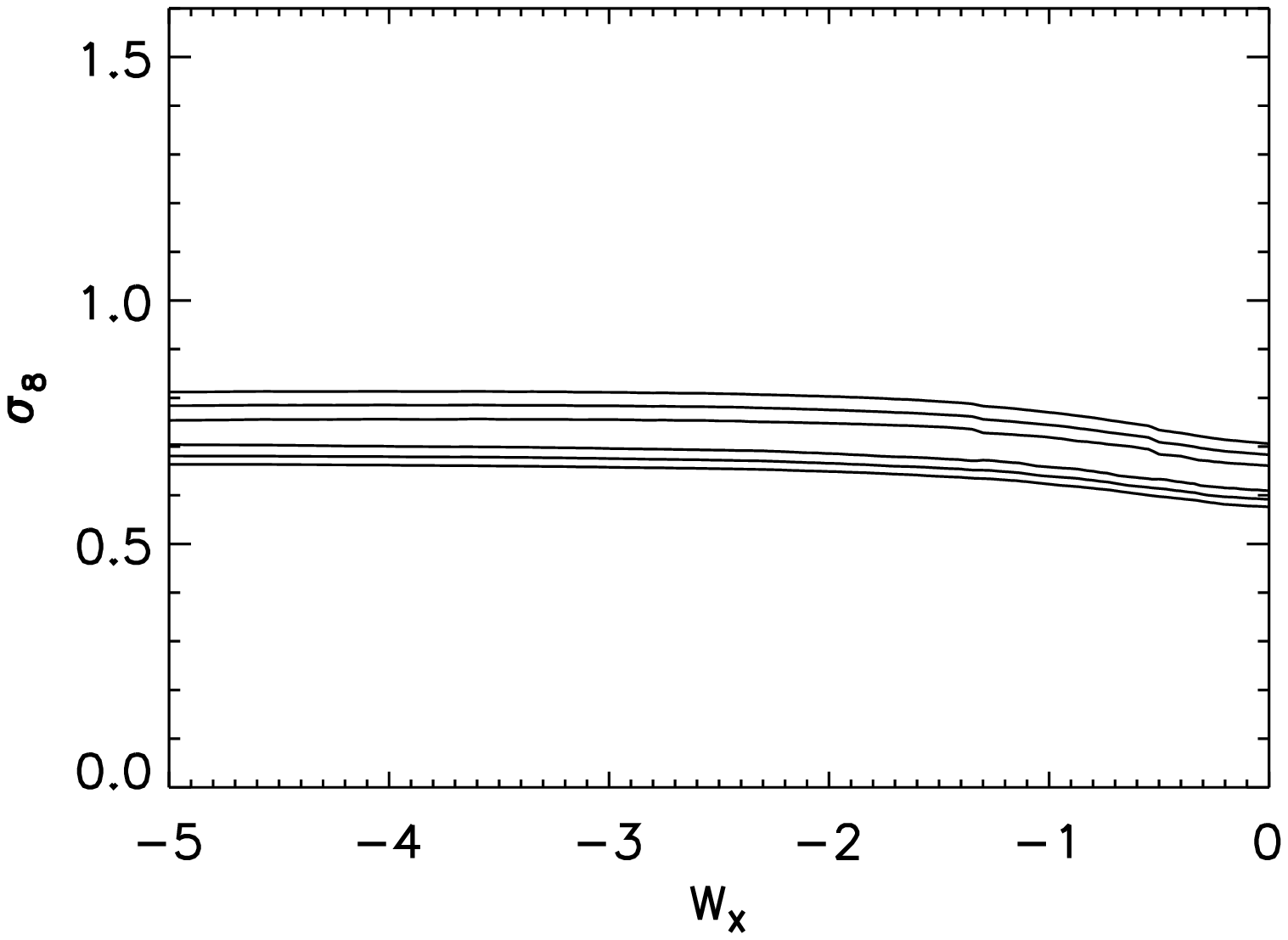,height=5.5cm,width=8.5cm}}
\vspace{-5.5cm}
\centerline{\hspace{ 8.2cm}
\psfig{figure=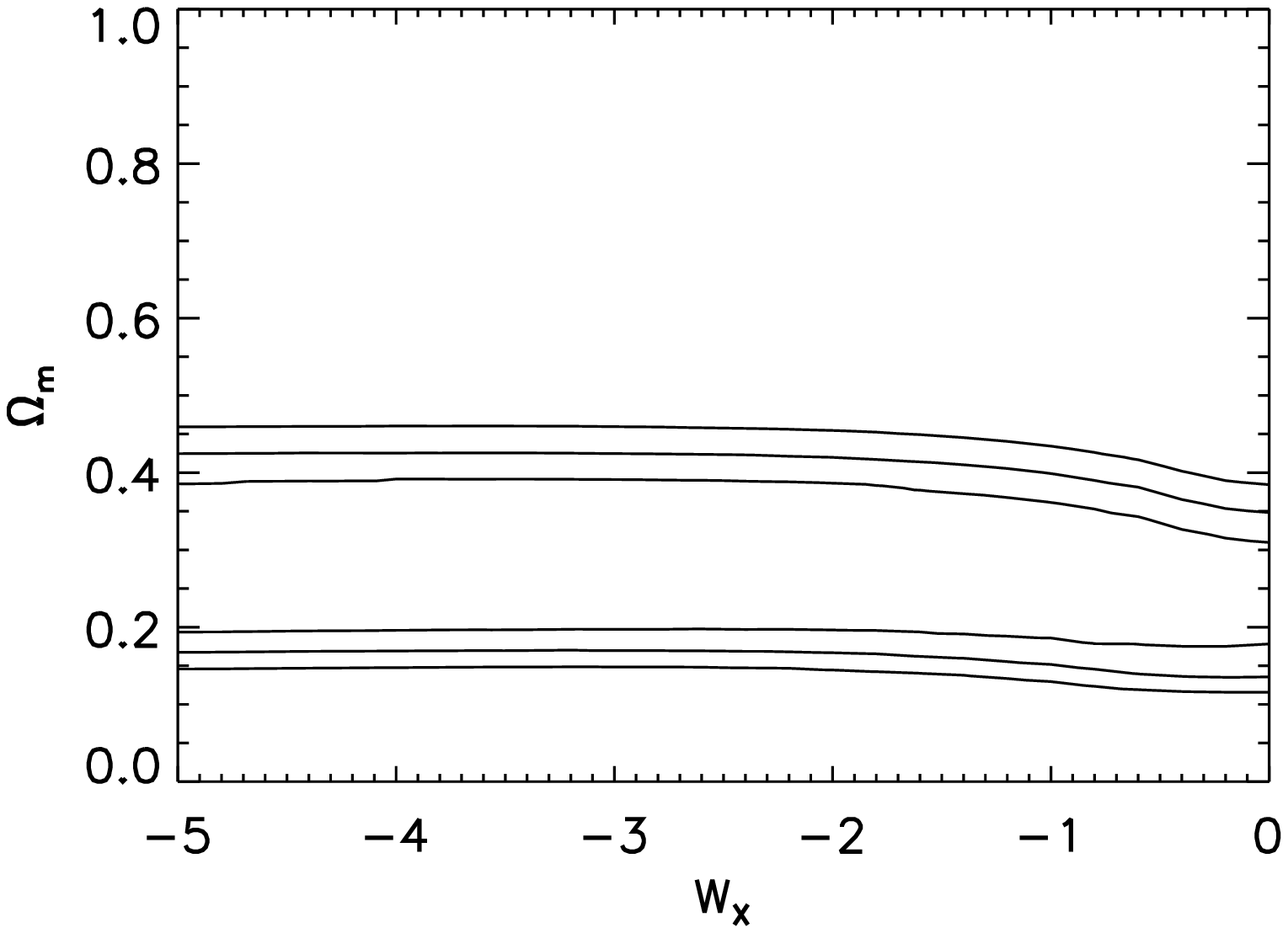,height=5.5cm,width=8.5cm}}
\vspace{-0.3cm}
\caption{\small Likelihood contours ($1$-$3\sigma$ levels for two
degrees of freedom) for various cosmological parameters obtained with
the abundances of the REFLEX clusters.  Note that the lower right
panel includes marginalization over $\sigma_8=[0.70,0.95]$. The
parameter priors of each diagram and the marginalization range are
given in the main text.}
\label{FIG_NZ}
\end{figure*}

This simple discussion shows that for the evaluation of NEC and SEC in
the given restrictive sense (\ref{NEC}, \ref{SEC}) one has to measure
$\Omega_{\rm m}$ and $w_{\rm x}$ (most importantly at the present
epoch). In order to get robust observational constraints we utilize
the complementarity of two approaches which have a different
sensitivity on $w_{\rm x}$ and $\Omega_{\rm m}$. Recent supernova (SN)
data constrain $w_{\rm x}$ and the cosmic matter density $\Omega_{\rm
m}$, but the results are highly degenerated (Garnavich et al. 1998,
Perlmutter et al. 1999).  The degeneracy can be broken by using
abundance measurements of a large sample of {\it nearby} X-ray
clusters of galaxies which give a precise estimate of $\Omega_{\rm m}$
(Schuecker et al. 2003) almost independent of $w_{\rm x}$ (see
Sect.\,\ref{REFLEX}). The results are compared with estimates obtained
by a complementary approach which combines SN, CMB and other data
(Sect.\,\ref{DISCUSS}).

\section{Observational data}\label{OBS}

\subsection{The X-ray cluster sample}\label{REFLEX}

The ROSAT ESO Flux-Limited X-ray (REFLEX) sample used for the present
investigations consists of the 452 X-ray brightest southern clusters
of galaxies with redshifts mainly below $z=0.3$. They are extracted
with a well-known selection function from the ROSAT All-Sky survey
(Voges et al. 1999) and confirmed by extensive optical follow-up
observations within a large ESO Key Programme (B\"ohringer et
al. 1998, Guzzo et al. 1999). The clusters are located in an area of
4.24\,sr in the southern hemisphere with Declination $\le 2.5$\,deg,
excluding galactic latitudes $|b|\le20$\,deg and some additional
crowded fields like the Magellanic Clouds (B\"ohringer et
al. 2001). The sample is expected to be at least $90\%$ complete. With
this sample the cluster X-ray luminosity function (B\"ohringer et
al. 2002), the spatial cluster-cluster correlation function (Collins
et al. 2000), its power spectrum (Schuecker et al. 2001), and the
cosmic matter density (Schuecker et al. 2002, 2003) have been
determined with unprecedented precision. The 426 REFLEX clusters used
for the present abundance measurements have at least 10 X-ray source
counts detected in the ROSAT energy band 0.5-2.0\,keV, X-ray
luminosities $L_{\rm X}\ge 2.5\times 10^{42}\,h^{-2}\,{\rm erg}\,{\rm
s}^{-1}$ and X-ray fluxes $S_{\rm X}\ge 3.0\times 10^{-12}\,{\rm
erg}\,{\rm s}^{-1}\,{\rm cm}^{-2}$ in the energy band
$0.1$-$2.4$\,keV.

The values of the cosmological parameters are estimated by comparing
the observed redshift histogram of the clusters with model
predictions. Note that the variances of the cosmological parameters
shown here are larger compared to those obtained with the combined
analysis of both the redshift histogram and the fluctuation power
spectrum as used in Schuecker et al. (2003). A brief description of
the model fits of the redshift histograms is given below. More details
especially the values of important model parameters can be found in
Schuecker et al. (2002, 2003). Deviations from the assumed values
introduce possible systematic errors leading to a comparatively large
scatter of, e.g., the measured $\sigma_8$ values (see Pierpaoli et
al. 2002), where $\sigma_8$ gives the standard deviation of the matter
density fluctuations in spheres with a comoving radius of
$8\,h^{-1}\,{\rm Mpc}$. In the present investigation we take these
systematic errors into account by marginalizing the cluster likelihood
distributions over a large $\sigma_8$ range (see below).

The number of clusters expected under REFLEX conditions at a specific
redshift is given by the integral over the mass function where the
lower mass limit is a function of redshift, flux-limit, cosmology etc.
The mass limit is directly related to an X-ray luminosity via the
mass/X-ray luminosity relation of galaxy clusters. We use an empirical
estimate given in Reiprich \& B\"ohringer (2002). The resulting mass
limit is transformed into the mass system defined in Jenkins et
al. (2001) so that the corresponding mass function can be integrated
to give the expected average number of clusters.

The computation of the X-ray luminosities takes into account the
systematic underestimation of the observed unabsorbed X-ray fluxes of
REFLEX clusters relative to the total fluxes (10\%, see B\"ohringer et
al. 2002) and the cosmic $K$-corrections obtained from a refined
Raymond-Smith code (B\"ohringer et al. 2000).  For the transformation
of the cluster masses defined at different overdensity radii the
Navarro et al. (1997) mass density profile is used with a redshift and
mass-independent concentration parameter of $c=5$. Deviations from
this value in the range $4\le c\le 6$ have effects below a few percent
and are neglected (Schuecker et al. 2003). We thus assume that the
REFLEX clusters do not show any significant evolution up to $z=0.3$ as
suggested by the redshift-independent distribution of the comoving
number densities of the REFLEX clusters (Schuecker et al. 2001, see
also the discussion in Rosati et al. 2002). The integration of the
mass function includes a convolution which takes into account the {\it
intrinsic} scatter of the mass/X-ray luminosity relation, and the
random flux (luminosity) errors of the REFLEX clusters as given in
B\"ohringer et al. (2001). In the present investigation we use an
effective scatter with a formal value of 25\,\% which includes the
contributions from flux errors (10\%) and intrinsic scatter
(20\%). The intrinsic scatter is in reasonable agreement with the
Reiprich \& B\"ohringer (2002) mass/X-ray luminosity relation if one
takes into account that realistic mass errors are expected to be a
factor 1.5 larger than the formal errors given in Reiprich \&
B\"ohringer (see Schuecker et al. 2003 for a more detailed
discussion).

The computation of the theoretical mass function assumes a matter
power spectrum and a model for the critical density contrast which
defines the virial cluster mass. For the matter power spectrum a Cold
Dark Matter transfer function with a given contribution of baryons is
assumed (see Eisenstein \& Hu 1998). No $w_{\rm x}$-dependent corrections of
the transfer functions as given in Ma et al. (1999) were applied
because of the comparatively small redshift range ($z<0.3$) and scale
range ($<1.5\,h^{-1}$\,Gpc) covered by the given X-ray cluster
sample. For the determination of the $w_{\rm x}$ (and $\Omega_{\rm m}$) dependent
critical density contrasts we follow the formalism of Wang \&
Steinhardt (1998) and integrate numerically the relevant ordinary
differential equations describing the collapse of a spherical
overdensity in an expanding Universe with $w_{\rm x}\neq-1$. We found that
the $\zeta$ function introduced in Wang \& Steinhardt (their equation
A11) used to compute the average critical density contrast between
cluster and background should be replaced by
\begin{equation}\label{WS1}
\zeta=5.4[\Omega_{\rm m}(z_{\rm ta})]^{-0.666} + 0.11
 [\Omega_{\rm m}(z_{\rm ta})]^{-1.85} (1-w_{\rm x})^{-1.5}\,,
\end{equation}
giving a good approximation (better than 5\%) for the range $-5<w_{\rm x}<0$
at turn-around ($z_{\rm ta}$). Eq.\,(\ref{WS1}) deviates from the
corresponding equation obtained for the smaller range $-1<w_{\rm x}<0$ given
in Wang \& Steinhardt by about 5\% and is larger by a factor 1.3 at
$w_{\rm x}=-5$ (for $\Omega_{\rm m}=0.3$). For completeness we also give the
equivalent linear overdensity of a virialized spherical shell,
\begin{equation}\label{WS2}
\delta_c\,=\, 1.686\, \Omega_{\rm m}^{0.037\, (1 - w_{\rm x})^{-2.7}}\,,
\end{equation}
which is needed when the Press-Schechter and the Sheth-Tormen mass
functions are used. In the present case we use the Jenkins et al. mass
function because of its higher precision where $\delta_c$ is not
needed. Nevertheless, for $-5<w<-1$ there is almost no change in
$\delta_c$ except for ridiculously small values of $\Omega_{\rm m}$.

The comparison of expected and observed cluster abundances assumes a
Gaussian likelihood distribution. For large sizes of the count cells
(as used here) this assumption is justified at a high level of
statistical significance by the measurements of Schuecker et
al. (2002, 2003).

The likelihood contours shown in Fig.\,\ref{FIG_NZ} illustrate the
sensitivity of the REFLEX sample to specific cosmological parameters
for a flat cosmic geometry. The set of default parameter values are:
$h=0.70$ (Freedman et al. 2001), $\sigma_8=0.711$ (normalization of
the matter power spectrum, see Schuecker et al. 2003, Allen et
al. 2002), $\Omega_{\rm m}=0.341$ (see Schuecker et al. 2003), $n_{\rm
S}=1.0$ (spectral index of initial scalar fluctuations, see the recent
CMB measurements given in Sect.\,\ref{INTRO}), $\Omega_{\rm b}h^2=0.022$
(baryon density, see CMB), $w_{\rm x}=-1$. Each panel in Fig.\,\ref{FIG_NZ}
shows the $1$-$3\sigma$ likelihood contours for two parameters whereby
the remaining parameter values are fixed by the default values given
above.

Notice that the cosmological parameters $\sigma_8$ and especially
$\Omega_{\rm m}$ can be obtained with nearby cluster samples almost
independent of the value of $w_{\rm x}$, as seen in the lower panels of
Fig.\,\ref{FIG_NZ}. $\Omega_{\rm m}$ measurements based on the abundance of
{\it nearby} clusters thus appear quite stable against the presence of
ordinary or more exotic $\Lambda$-like energies.

The lower right panel of Fig.\,\ref{FIG_NZ} gives the final
$(w_{\rm x},\Omega_{\rm m})$ likelihood values which will be used for the
combination with the SN data (see Sect.\,\ref{RESULTS}). The
likelihood distribution is obtained with $h=0.70$, $n_{\rm S}=1$,
$\Omega_{\rm b}h^2=0.022$ and after marginalization over the quite large
$\sigma_8$ range of $[0.70,0.95]$. This interval covers most of the
$\sigma_8$ values obtained recently with different samples, model
assumptions, and methods (weak lensing, optical clusters, X-ray
cluster temperature and luminosity functions, power spectrum,
Sunyaev-Zel'dovich effect power spectrum, galaxy clustering etc.) as
summarized in, e.g., Pierpaoli et al. (2002). Compared to this
marginalization range, the relatively small systematics introduced by
the uncertainties in the values of $n_{\rm S}$, $h$, and $\Omega_{\rm b}h^2$
are neglected (see Fig.\,\ref{FIG_NZ}, see also Table\,1 in Schuecker
et al. 2003).

\subsection{The type-Ia supernova samples}\label{SNIA}

\begin{figure*}
\vspace{-0.0cm}
\centerline{\hspace{-10.5cm}
\psfig{figure=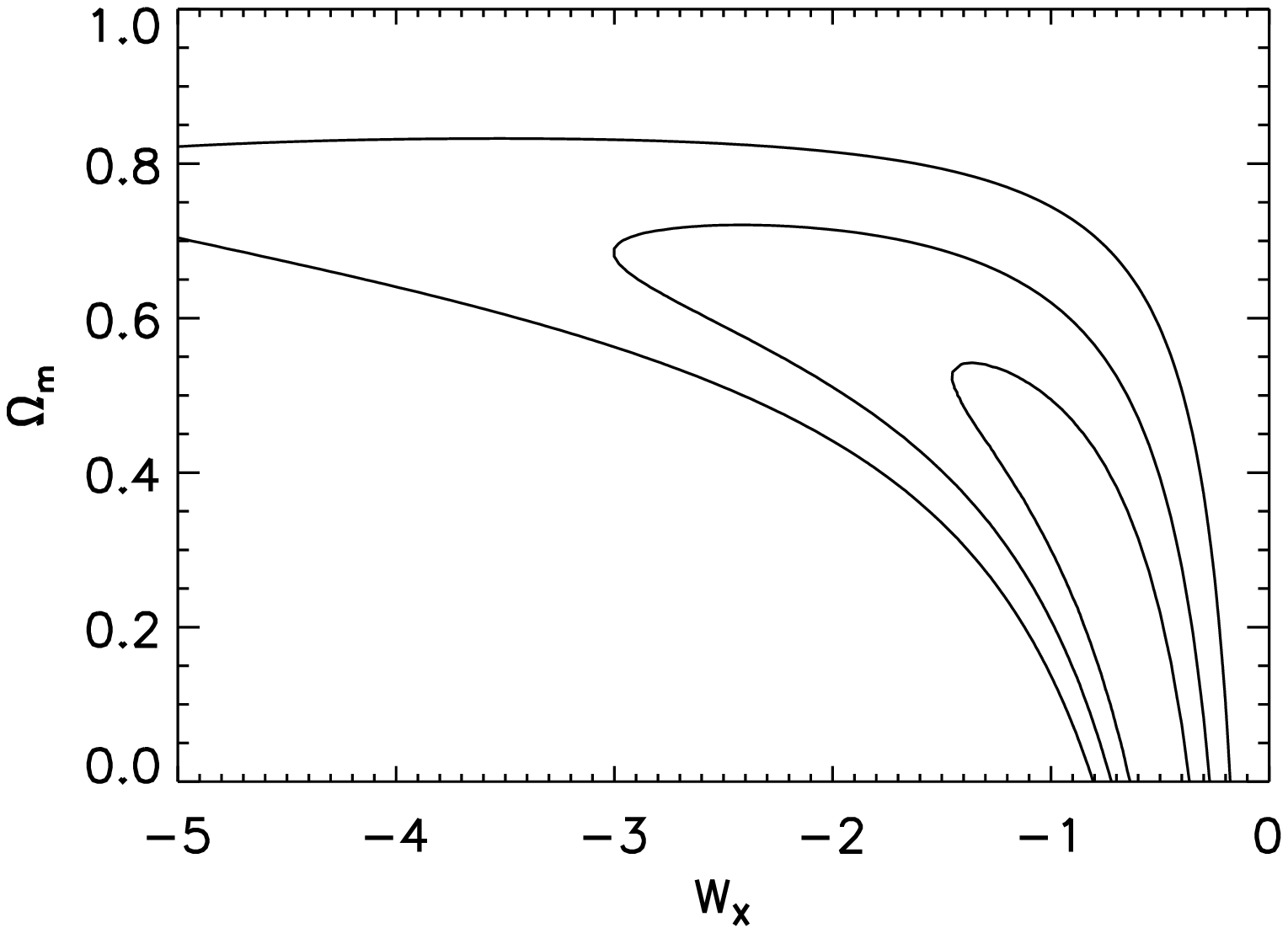,height=5.5cm,width=8.5cm}}
\vspace{-5.5cm}
\centerline{\hspace{ 8.2cm}
\psfig{figure=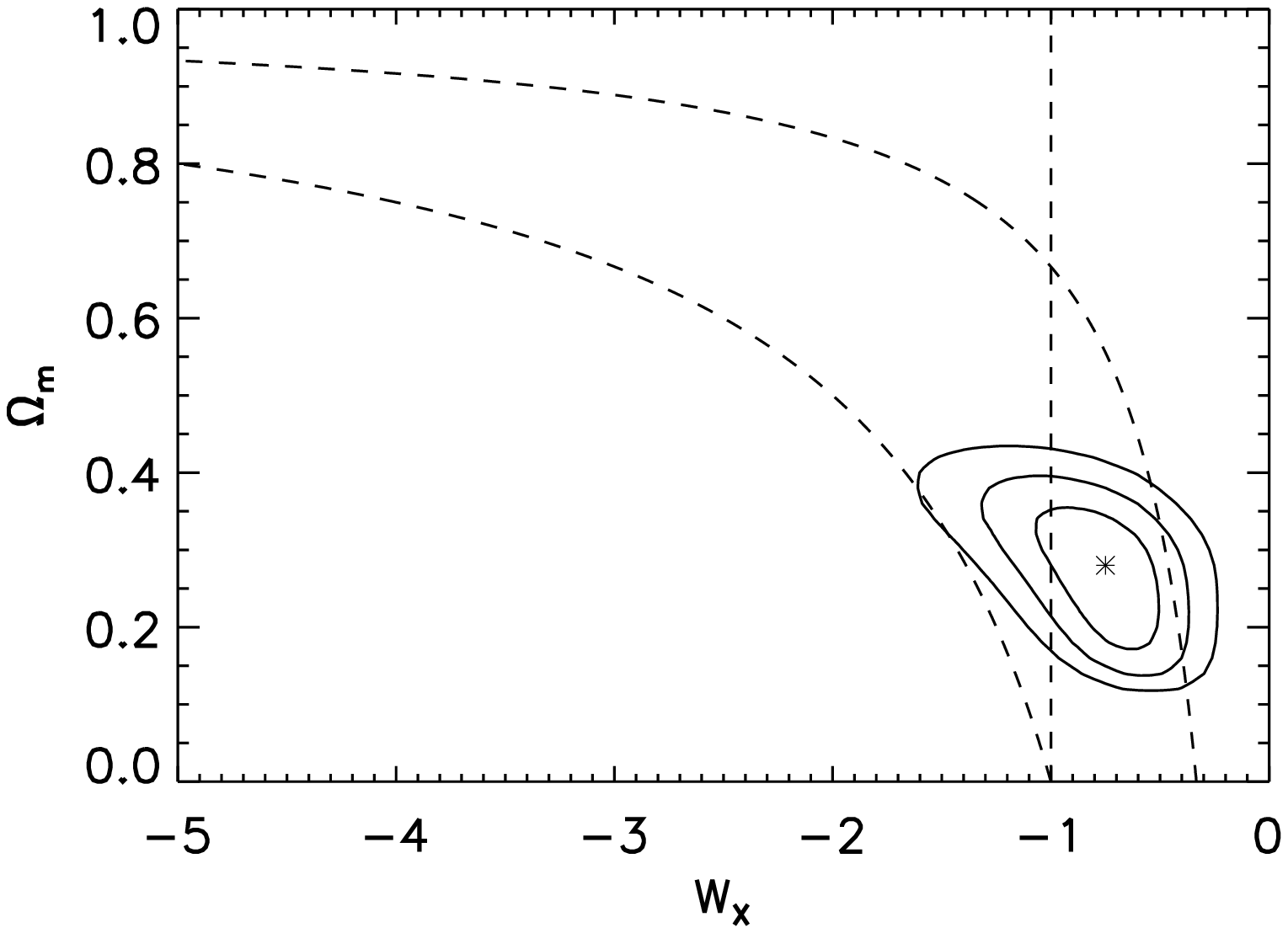,height=5.5cm,width=8.5cm}}
\vspace{-0.0cm}
\centerline{\hspace{-10.5cm}
\psfig{figure=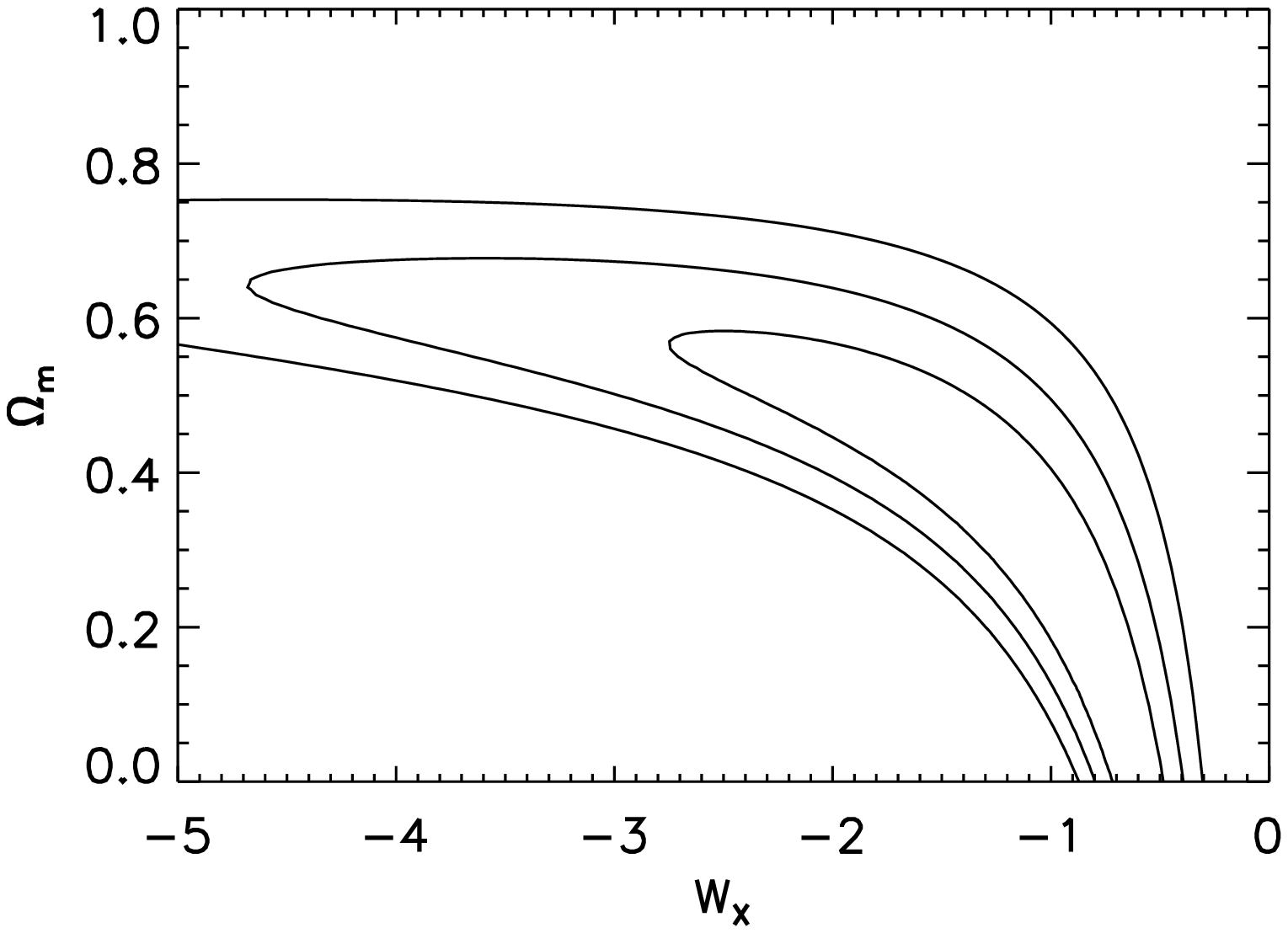,height=5.5cm,width=8.5cm}}
\vspace{-5.5cm}
\centerline{\hspace{ 8.2cm}
\psfig{figure=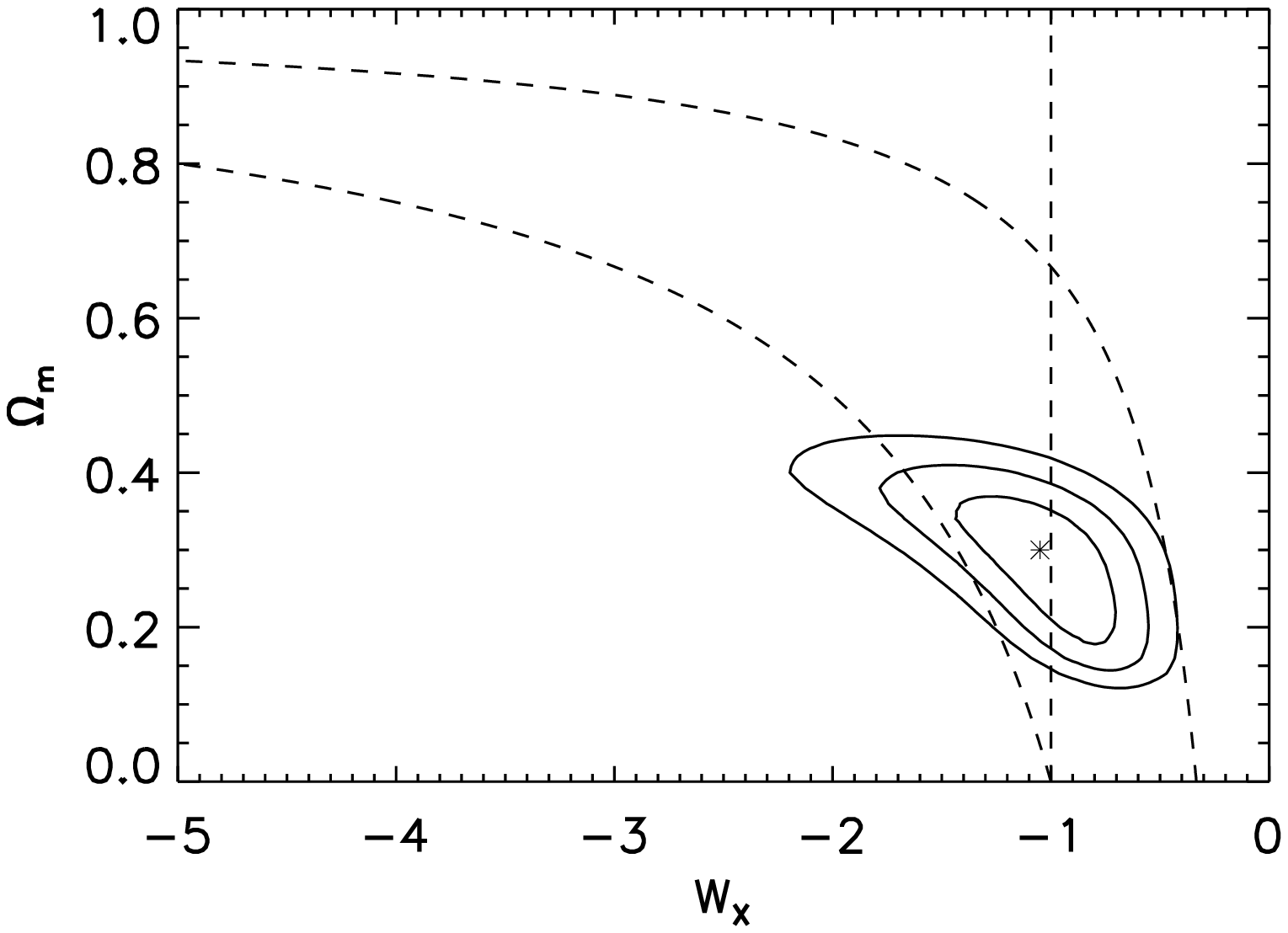,height=5.5cm,width=8.5cm}}
\vspace{-0.0cm}
\centerline{\hspace{-10.5cm}
\psfig{figure=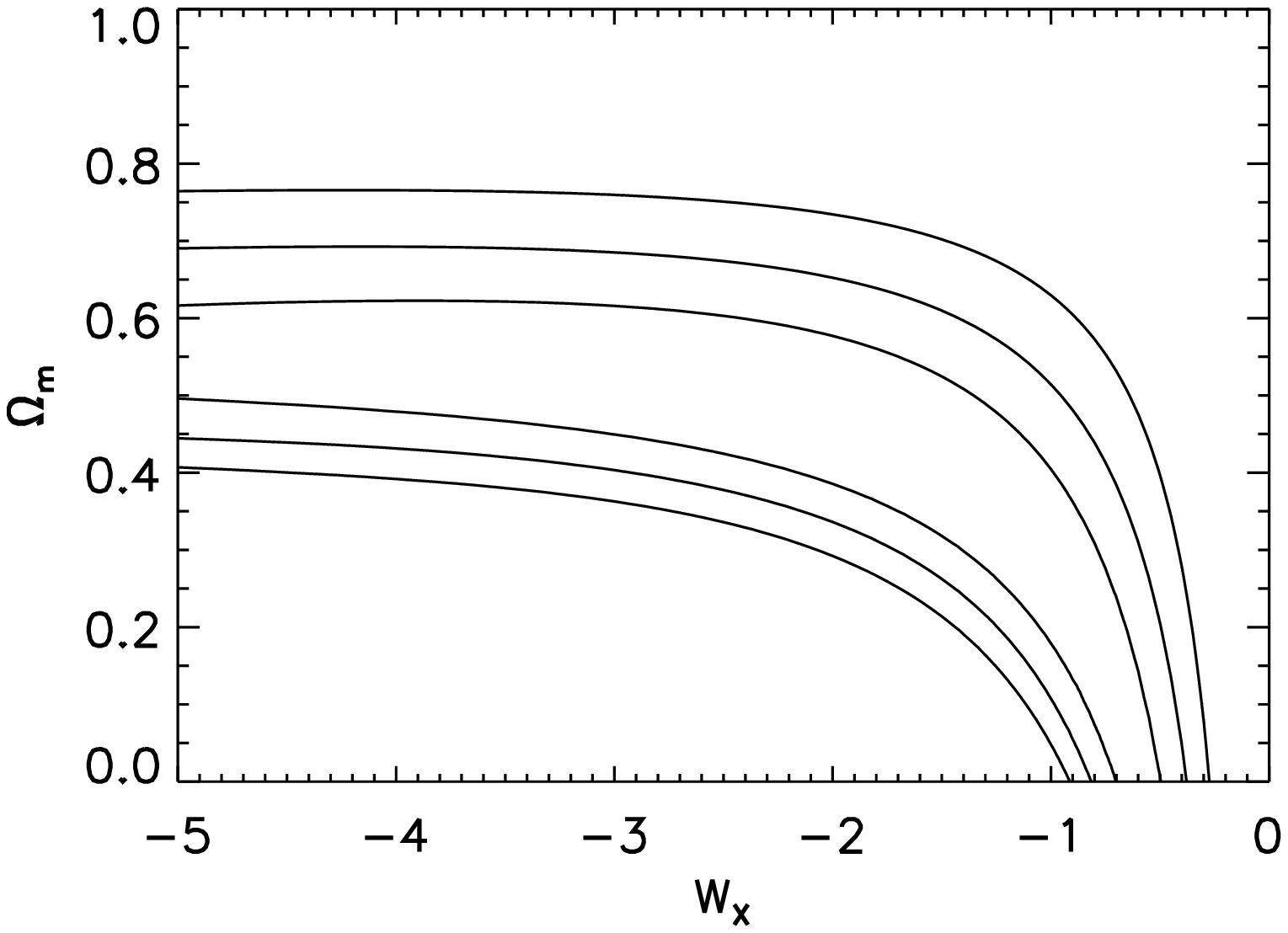,height=5.5cm,width=8.5cm}}
\vspace{-5.5cm}
\centerline{\hspace{ 8.2cm}
\psfig{figure=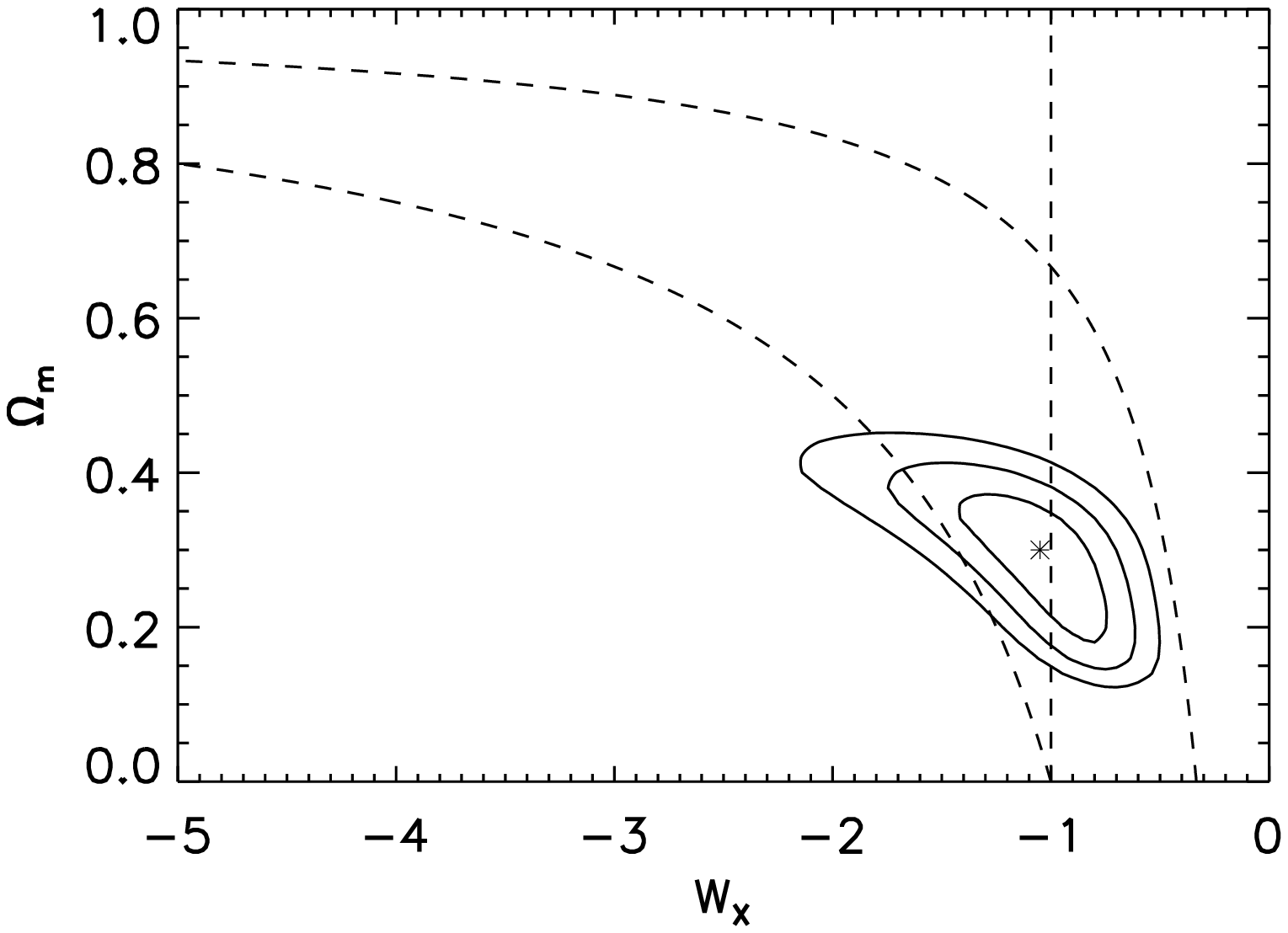,height=5.5cm,width=8.5cm}}
\vspace{-0.3cm}
\caption{\small Likelihood contours ($1$-$3\sigma$ levels for two
degrees of freedom) obtained with SNe\,Ia only (left panels) and with
SNe\,Ia plus REFLEX X-ray clusters (right panels) marginalized over
$\sigma_8=[0.70,0.95]$. The SN results of the first two rows are based
on the Riess et al. (1998) sample (MCLS corrections upper row, $\Delta
m_{15}$ corrections middle row), whereas the data in the lower row are
based on the Perlmutter et al. (1999) sample. Vertical dashed lines at
$w_{\rm x}=-1$ represent models with Einstein's cosmological
constant. Curved dashed lines devide the parameter spaces into sectors
where SEC is valid indicating a decelerated cosmic expansion (above
the upper dashed curves), sectors where SEC is violated but NEC is
fulfilled indicating an accelerated cosmic expansion (between the
dashed curves), and sectors where NEC (and thus SEC) are violated
indicating a super-accelerated cosmic expansion (below the lower
dashed curves).}
\label{FIG_NZSN1}
\end{figure*}

The two SNe\,Ia samples described in Riess et al. (1998) and
Perlmutter et al. (1999) are used for the present investigations. The
distance moduli of the SNe are determined with spectral and
photometric observations.  The host-galaxy subtracted SN peak
magnitudes are corrected by the two teams for the cosmic $K$-effect,
absorption in the Galaxy and host galaxy assuming a standard galactic
absorption law and no colour evolution (Perlmutter et al. did not
correct for absorption in the host galaxy but made some additional
checks and rejected obviously reddened SNe from the cosmological
tests), time dilation in the light curve, and the shape of the light
curve.  There is a distinct difference in the treatment of the latter
correction in the two samples which complicates a direct comparison
between the two data sets. The difference is, however, not relevant
for cosmological applications.

Corrections of the change in the peak absolute luminosity are
performed for each SN\,Ia in the Riess et al. sample with the $\Delta
m_{15}$ method of Phillips (1993), Hamuy et al. (1995), and Phillips
et al. (1999), and with the Multi-Color Light Curve Shape (MCLS)
method of Riess et al. (1998).  A simple re-normalization of the
observed apparent peak magnitude is performed for each SN\,Ia in the
Perlmutter et al. sample using the stretch factor introduced in
Perlmutter et al. (1995, 1997).

The Riess et al. sample consists of 27 nearby SNe\,Ia ($z<0.2$), 10
High-$z$ SNe\,Ia ($0.30\le z\le 0.97$) and 6 High-$z$ Snapshot SNe\,Ia
($0.16\le z \le 0.83$). The resulting catalogue used here gives the
redshifts and the distance moduli as obtained with the $\Delta m_{15}$
and with the MCLS methods. The magnitudes are corrected for the
various effects mentioned above. The errors of the distance moduli and
cosmological redshifts of the SNe are used for weighting.

The Perlmutter et al. sample consists of 38 SCP SNe\,Ia ($0.172\le z
\le 0.830$) and 16 Cal\'{a}n/Tololo SNe\,Ia ($0.014\le z \le
0.101$). The sample is the same as used by Perlmutter et al. for their
Primary Fit (C-fit). Note that this sample excludes 6 SNe from the
original sample of 60 SNe (2 residual outliers, 2 stretch outliers, 2
likely reddened SNe). The resulting catalogue used here gives the SN
redshifts, the effective (corrected) peak apparent magnitudes and the
total uncertainty of the magnitudes. The latter quantities already
include the uncertainties related to the expected errors of the
cosmological redshifts of the SNe. 

Note that the 16 low redshift Cal\'{a}n/Tololo SNe\,Ia and two distant
SNe are members of both SN samples. Therefore, we cannot regard the
Riess et al. and the Perlmutter et al. SN sample as statistically
independent.

The present SN likelihood analyses assume Gaussian likelihood
functions as verified in Riess et al. and Perlmutter et al.  With the
corrected peak magnitudes and redshifts of the SNe it is
straightforward to compare observed and model magnitudes assuming
different values of $w_{\rm x}$ and $\Omega_{\rm m}$ (including marginalization
over method-specific quantities as in Riess et al. and Perlmutter et
al.). The results are shown for the different samples and light curve
corrections in the left panels of Fig.\,\ref{FIG_NZSN1}.

\begin{figure*}
\vspace{-0.0cm}
\centerline{\hspace{-12.0cm}
\psfig{figure=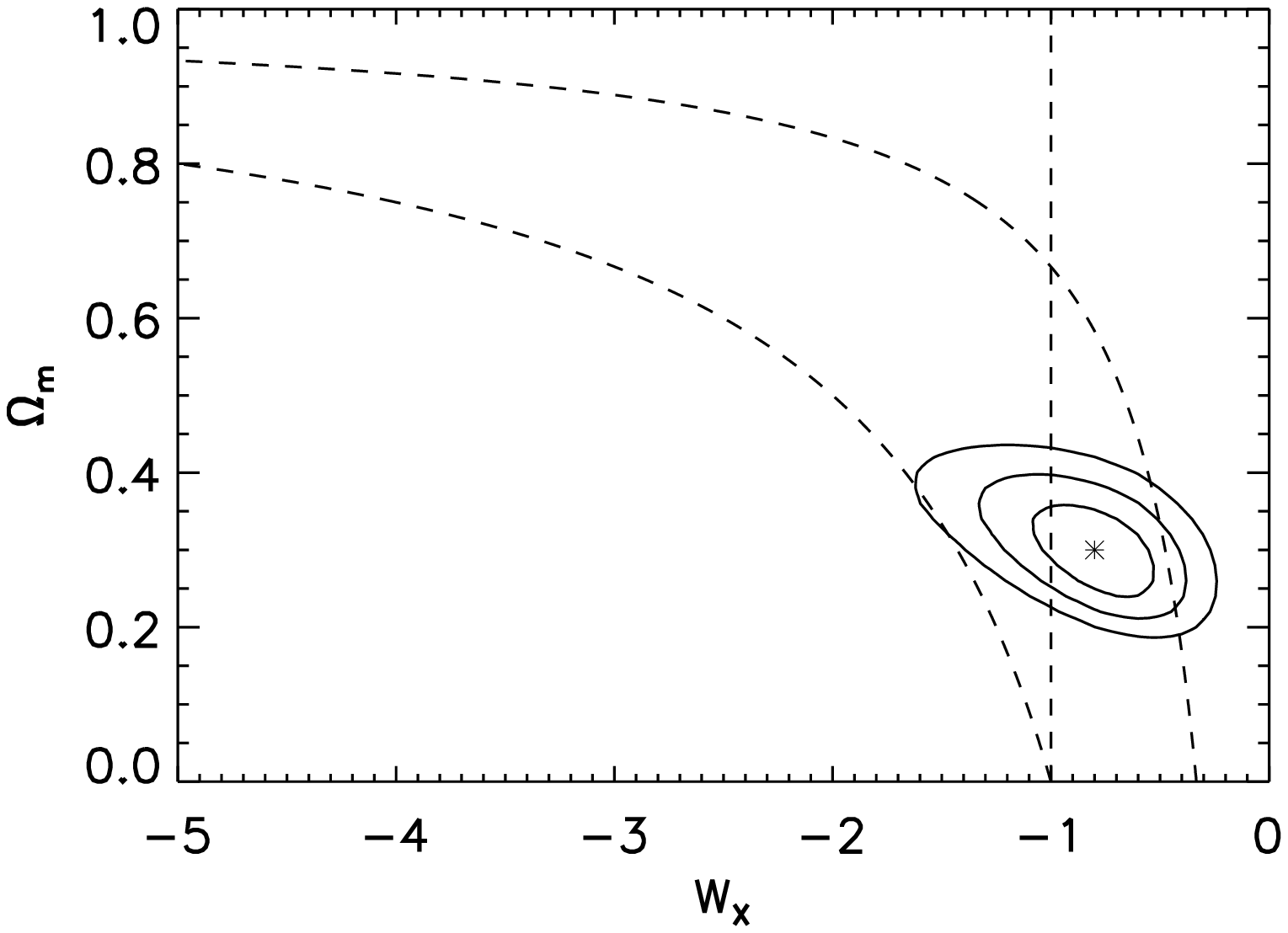,height=4.5cm,width=6.0cm}}
\vspace{-4.5cm}
\centerline{\hspace{-1.0cm}
\psfig{figure=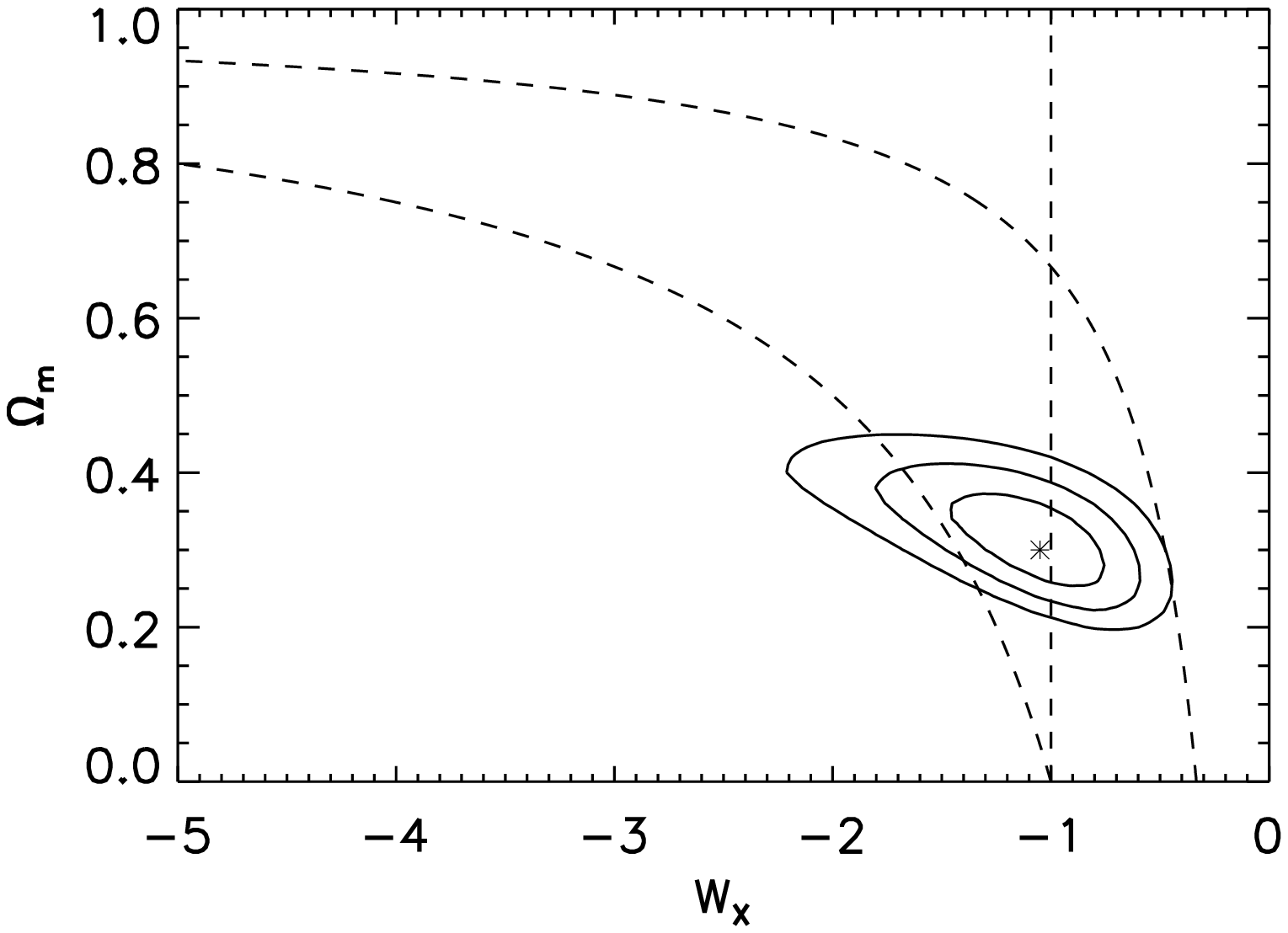,height=4.5cm,width=6.0cm}}
\vspace{-4.5cm}
\centerline{\hspace{10.0cm}
\psfig{figure=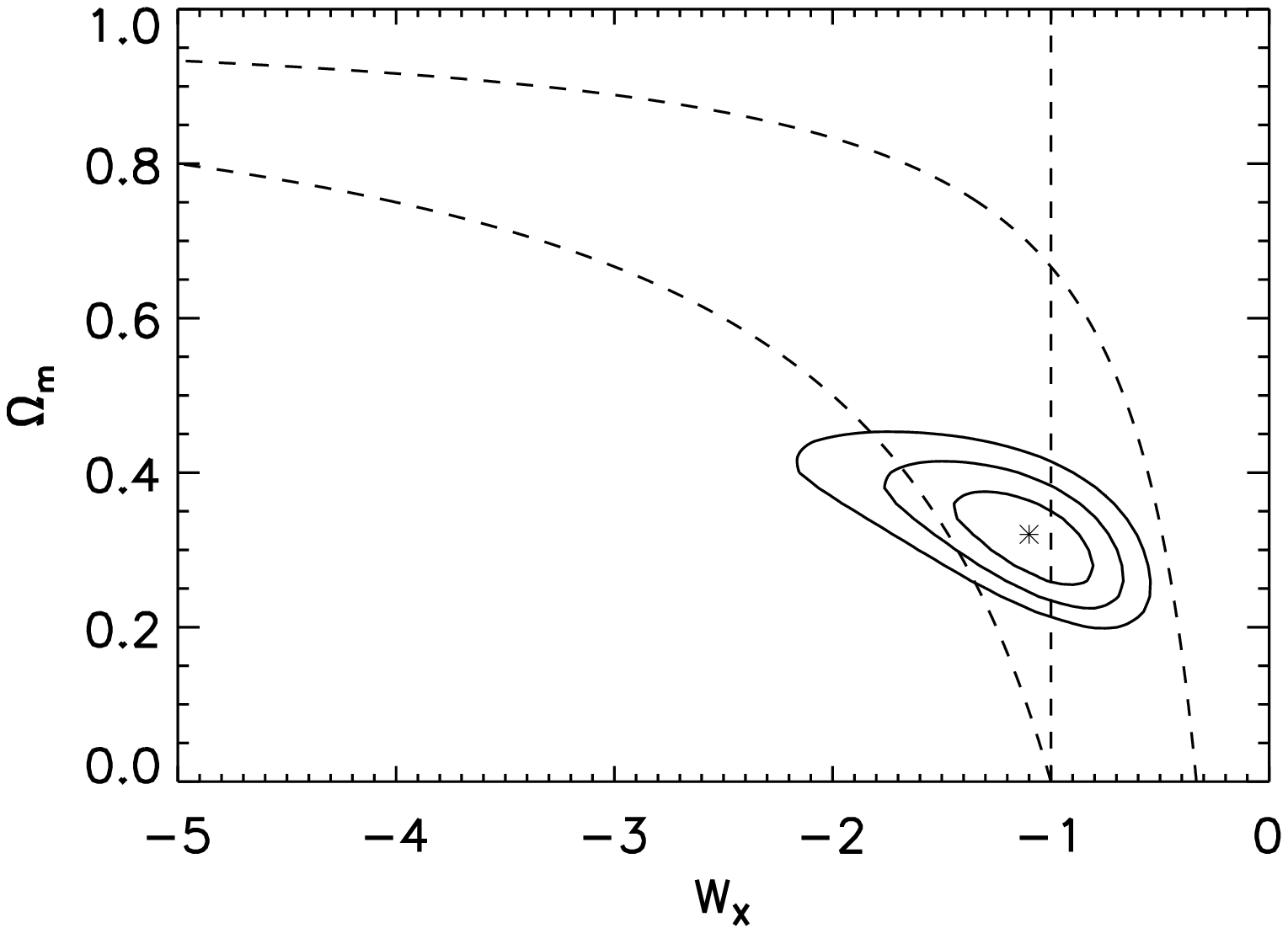,height=4.5cm,width=6.0cm}}
\vspace{-0.0cm}
\centerline{\hspace{-12.0cm}
\psfig{figure=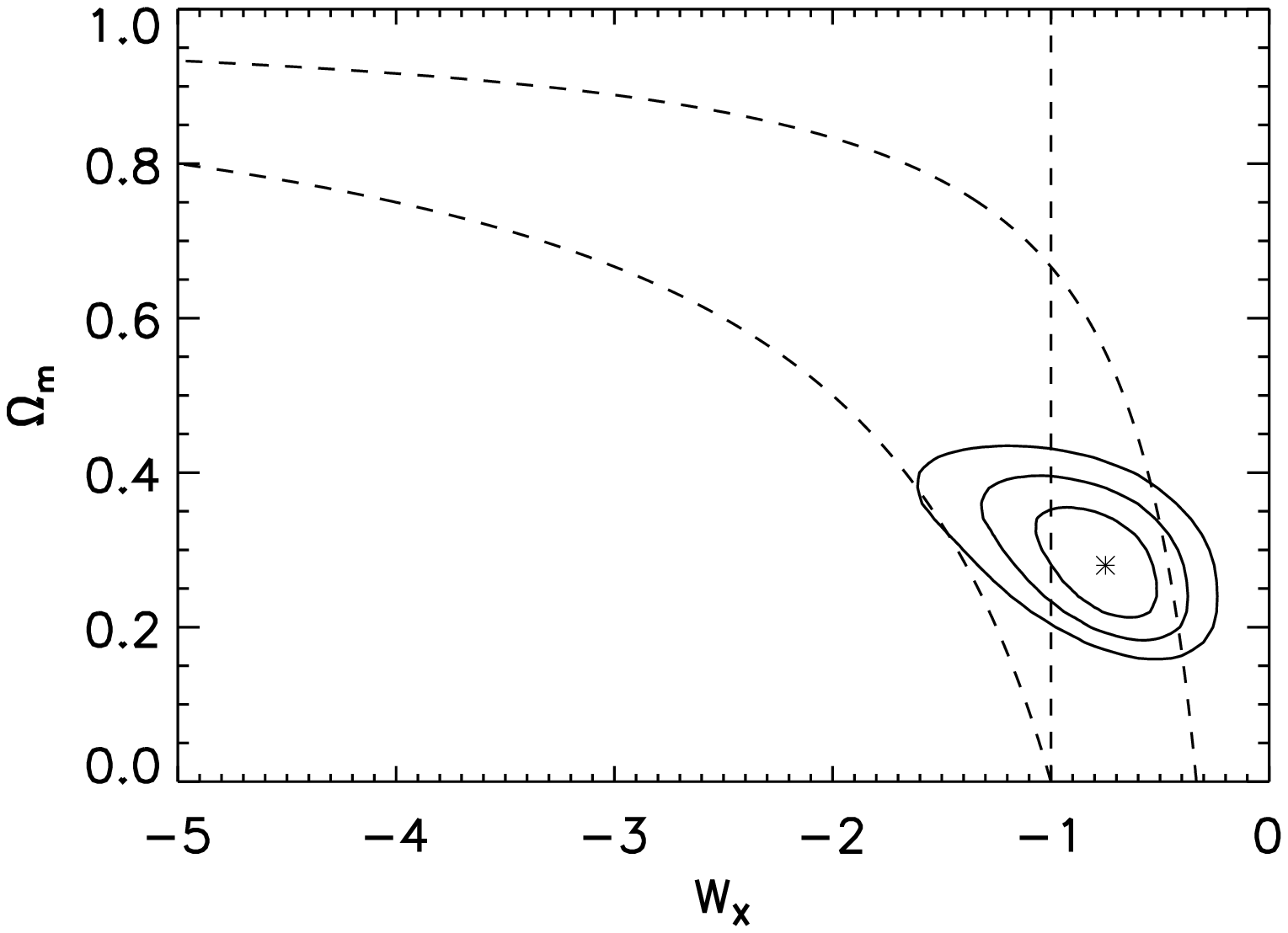,height=4.5cm,width=6.0cm}}
\vspace{-4.5cm}
\centerline{\hspace{-1.0cm}
\psfig{figure=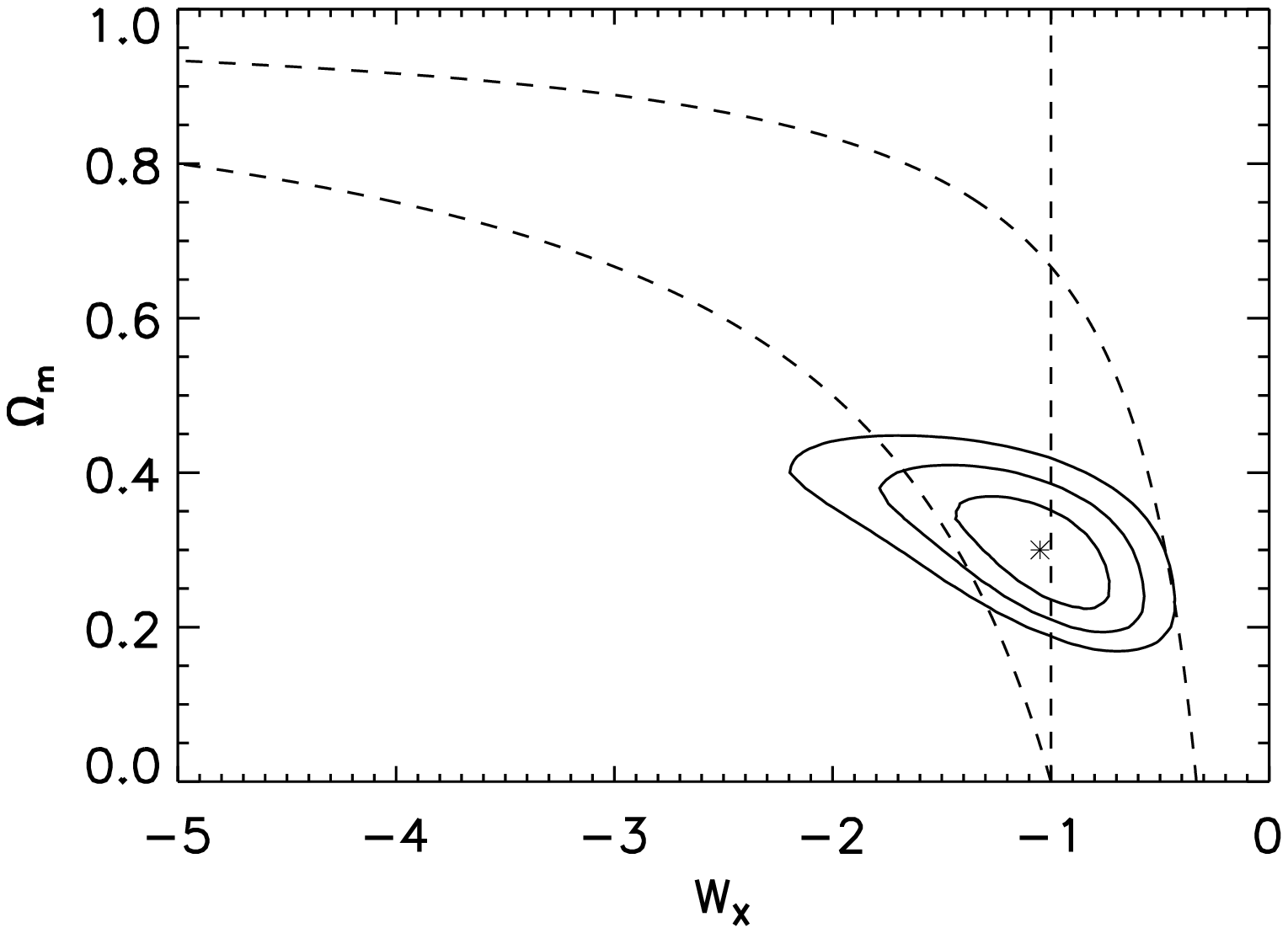,height=4.5cm,width=6.0cm}}
\vspace{-4.5cm}
\centerline{\hspace{10.0cm}
\psfig{figure=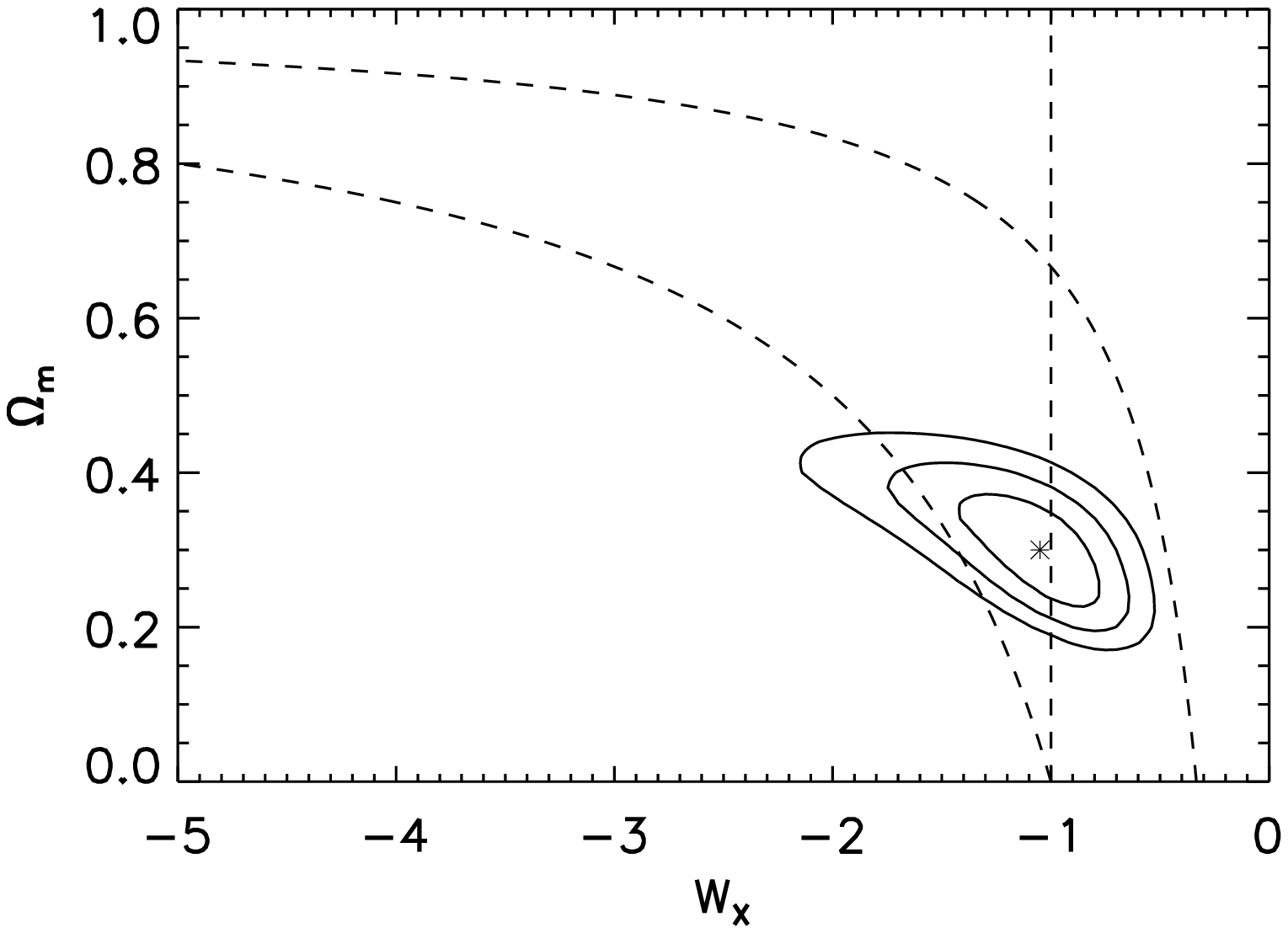,height=4.5cm,width=6.0cm}}
\vspace{-0.0cm}
\centerline{\hspace{-12.0cm}
\psfig{figure=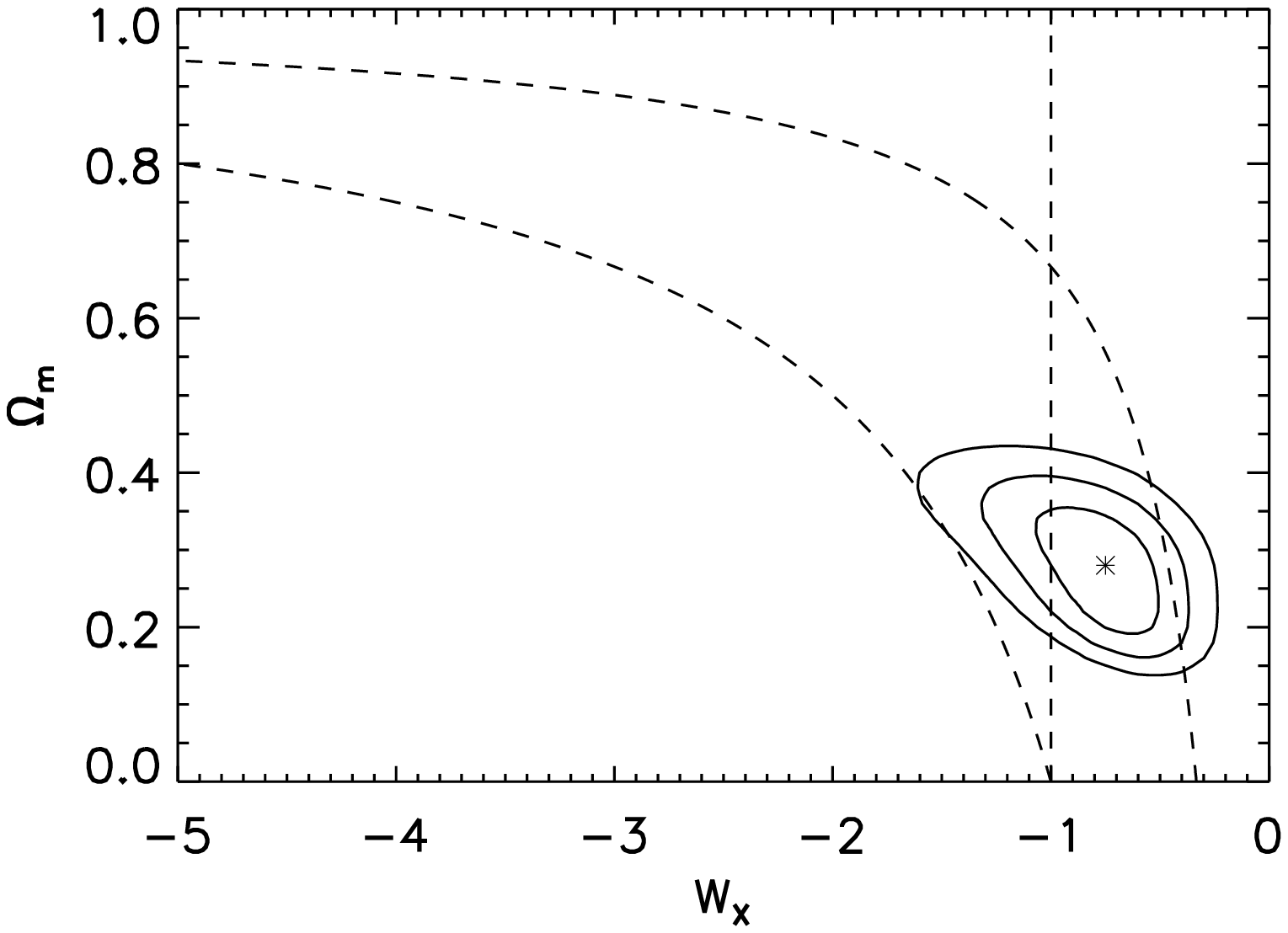,height=4.5cm,width=6.0cm}}
\vspace{-4.5cm}
\centerline{\hspace{-1.0cm}
\psfig{figure=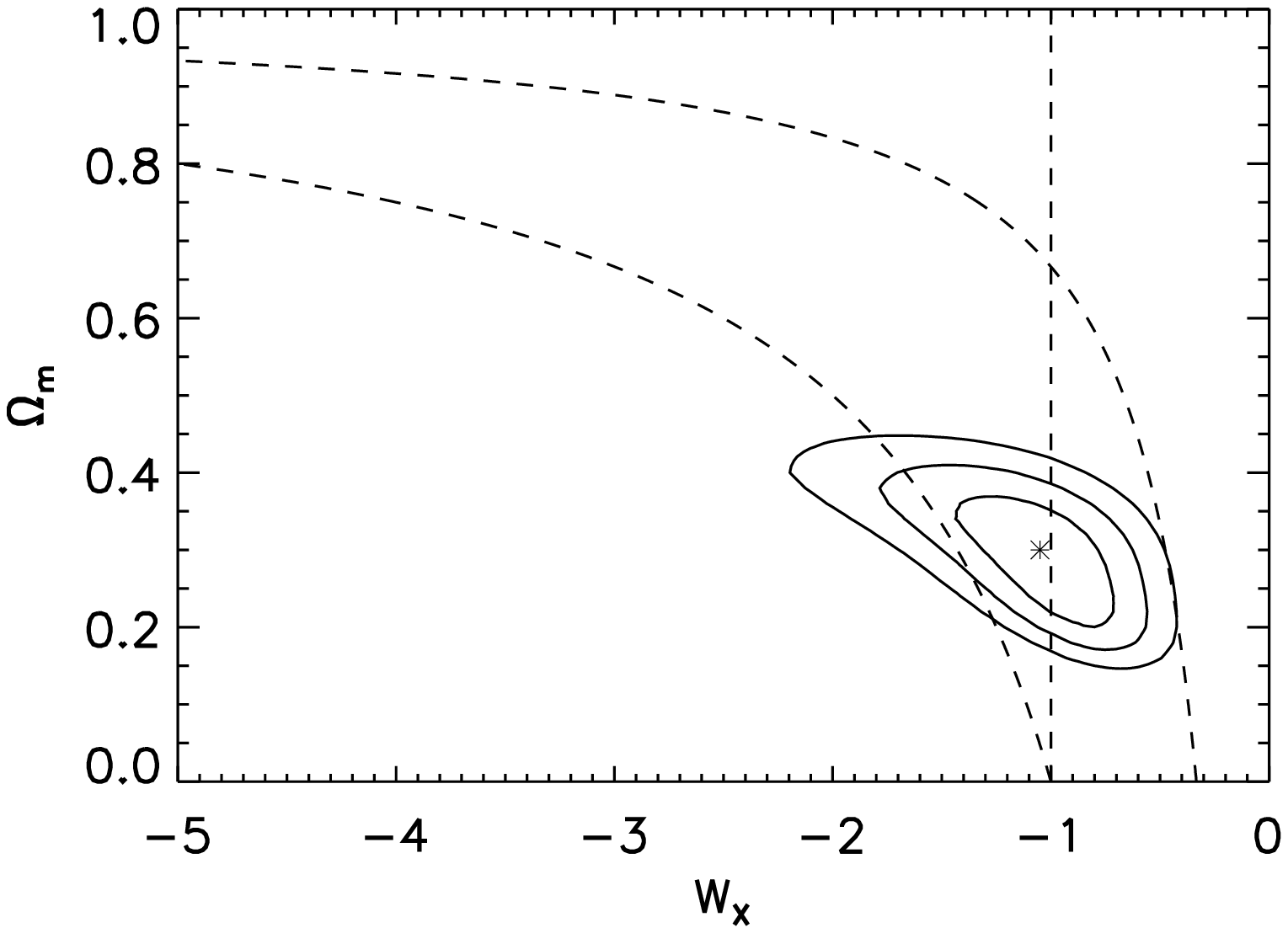,height=4.5cm,width=6.0cm}}
\vspace{-4.5cm}
\centerline{\hspace{10.0cm}
\psfig{figure=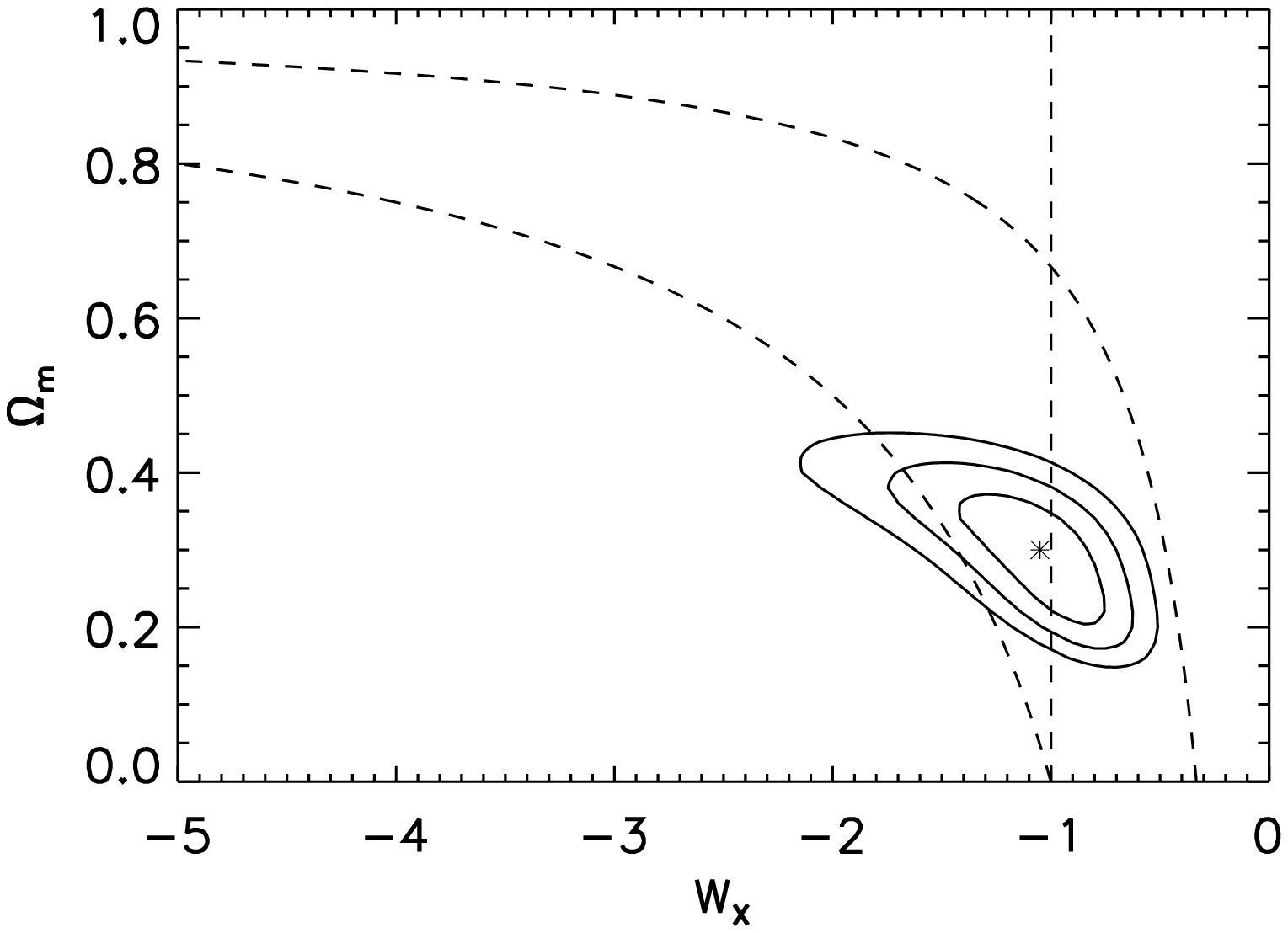,height=4.5cm,width=6.0cm}}
\vspace{-0.0cm}
\centerline{\hspace{-12.0cm}
\psfig{figure=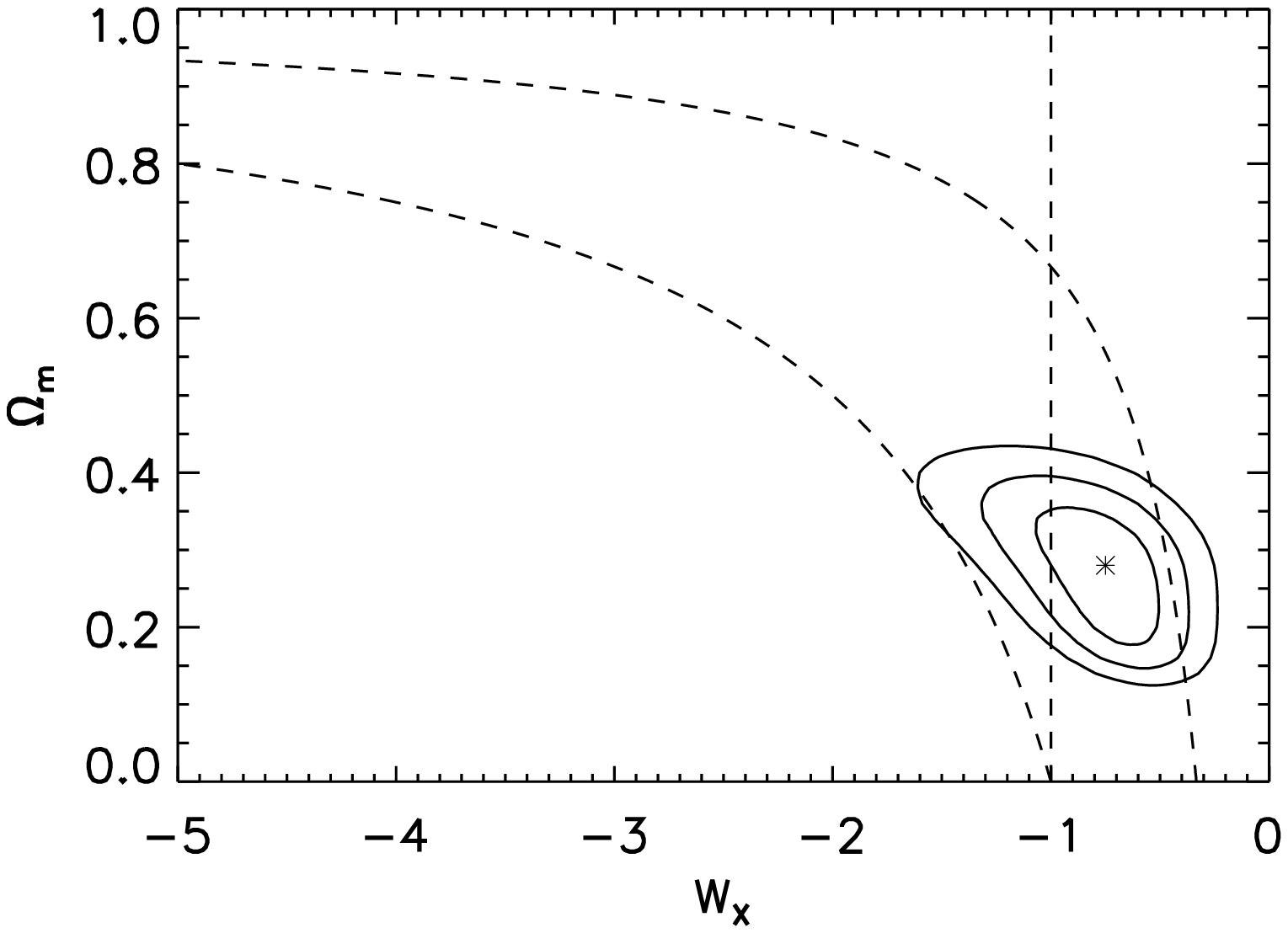,height=4.5cm,width=6.0cm}}
\vspace{-4.5cm}
\centerline{\hspace{-1.0cm}
\psfig{figure=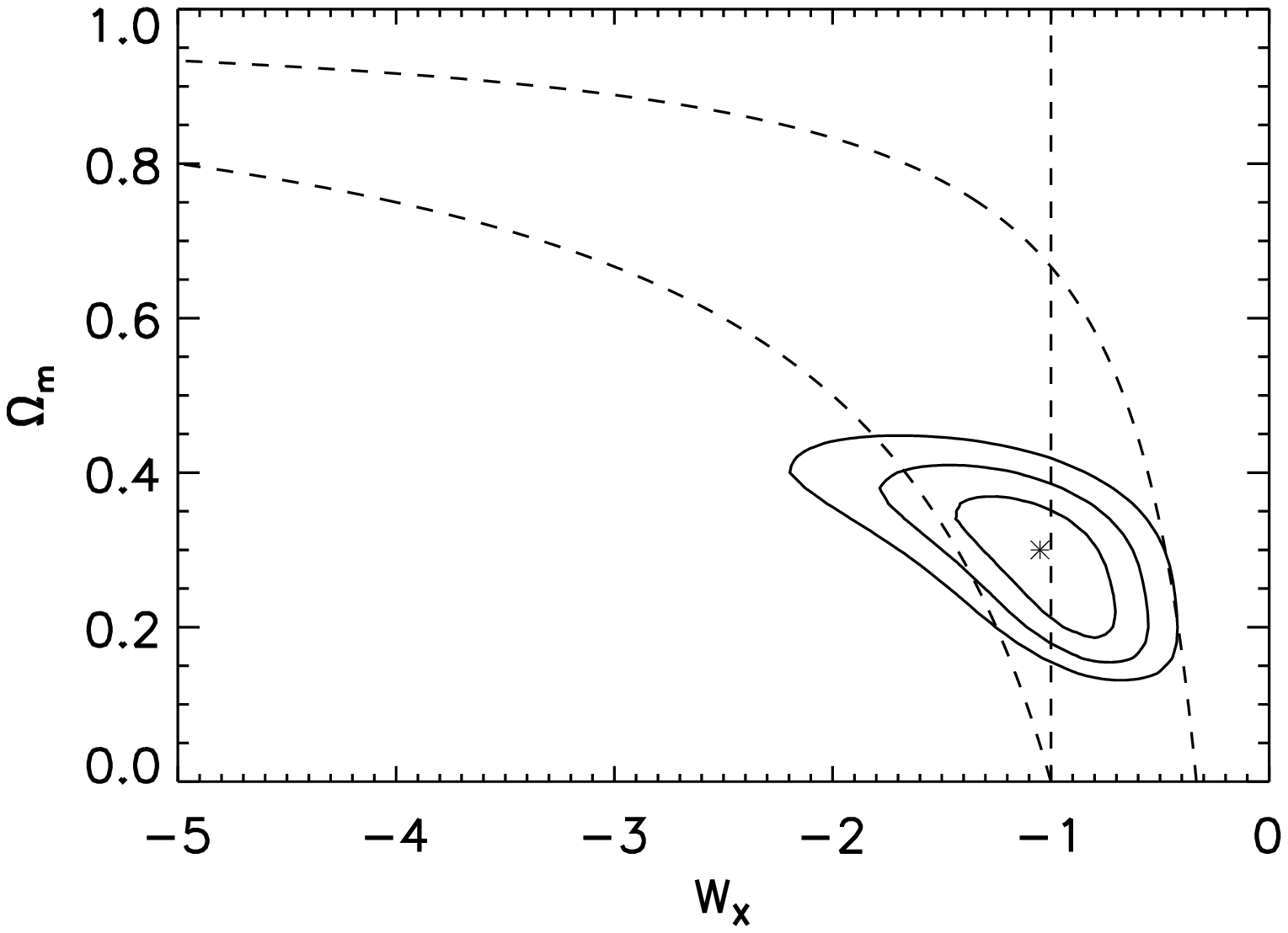,height=4.5cm,width=6.0cm}}
\vspace{-4.5cm}
\centerline{\hspace{10.0cm}
\psfig{figure=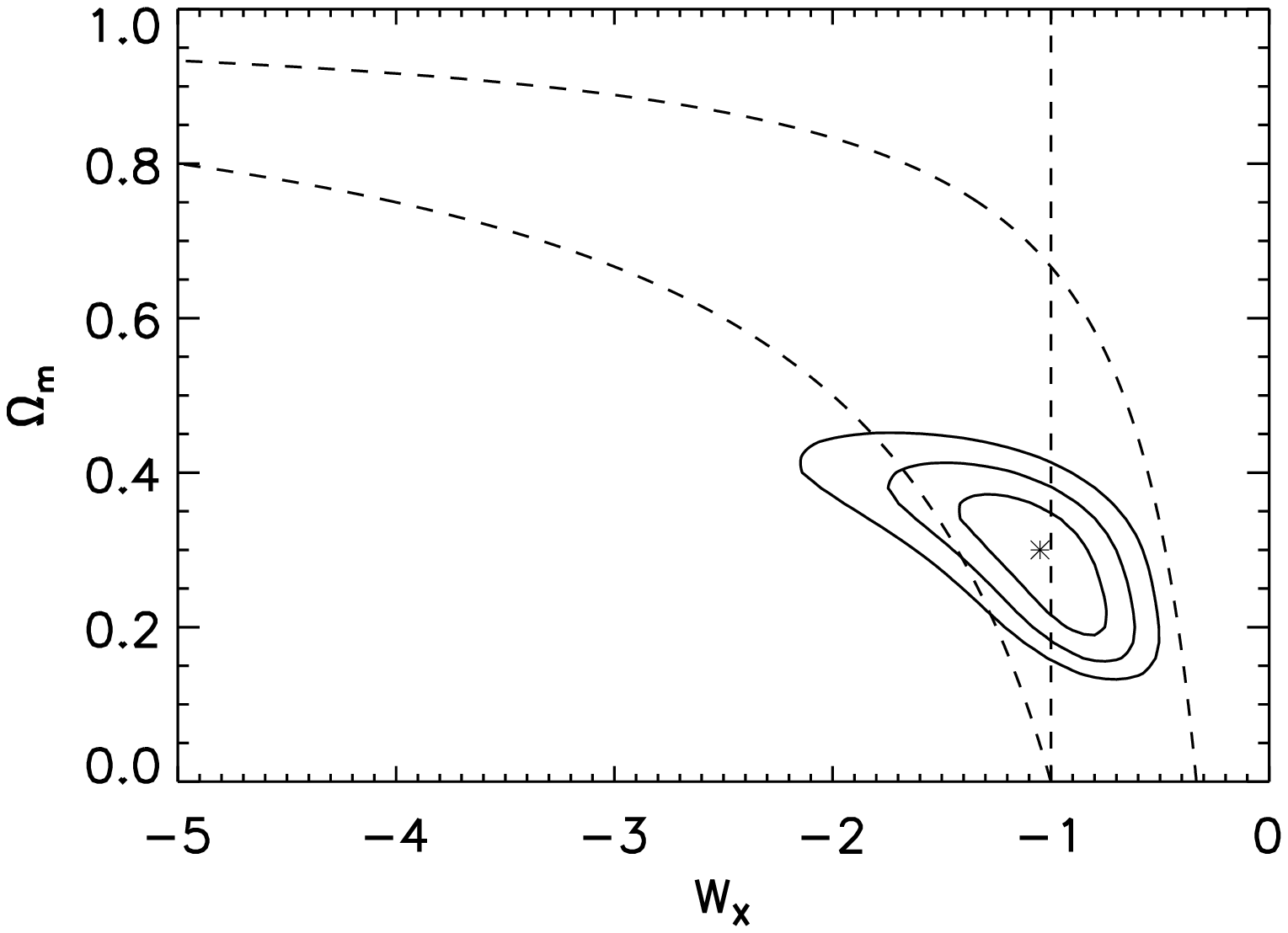,height=4.5cm,width=6.0cm}}
\vspace{-0.0cm}
\caption{\small Likelihood contours ($1$-$3\sigma$ levels for two
degrees of freedom) obtained with SNe\,Ia plus REFLEX X-ray clusters
illustrating the stability of the SEC and NEC tests by changing the
$\sigma_8$ ranges for marginalization. First column: Riess et
al. (MCLS corrections) plus REFLEX. Second column: Riess et
al. ($\Delta m_{15}$ corrections) plus REFLEX. Third column:
Perlmutter et al. plus REFLEX. First row:
$\sigma_8=[0.70,0.75]$. Second row: $\sigma_8=[0.70,0.80]$, Third row:
$\sigma_8=[0.70,0.85]$, Fourth row: $\sigma_8=[0.70,0.90]$.}
\label{FIG_NZSN2}
\end{figure*}

\section{Combined constraints from X-ray clusters and SNe\,-Ia}\label{RESULTS}

The combination of the likelihood distributions obtained with the
X-ray clusters and SNe makes the realistic assumption that both
samples are statistically independent so that the cluster and SN
likelihoods can be point-wise multiplicated. The resulting joint
likelihood distributions of the constraints on $w_{\rm x}$ and $\Omega_{\rm m}$
obtained for the three SN samples combined with the REFLEX results are
shown in the right panels of Fig.\,\ref{FIG_NZSN1} (see also
Table\,1). As discussed in Sect.\,\ref{TEST}, vertical dashed lines at
$w_{\rm x}=-1$ represent the case of a cosmological constant and devide the
parameter spaces into the dark energy sectors with $-1<w_{\rm x}<0$ and the
phantom energy sectors with $w_{\rm x}<-1$. The curved dashed lines are
computed with Eqs.\,(\ref{NEC}) and (\ref{SEC}) and give the deviding
lines for NEC (lower dashed curves) and SEC (upper dashed curves).

Figure\,\ref{FIG_NZSN1} shows that for all three combinations of X-ray
cluster data obtained with the marginalization interval
$\sigma_8=[0.70,0.95]$ and SN data the centroids of the joint
likelihood distributions range between $-1.05\le w_{\rm x}\le -0.75$ and
$0.28\le \Omega_{\rm m}\le 0.30$ (see also Table\,1).

Figure\,\ref{FIG_NZSN2} illustrates the stability of the results by
showing the combined likelihood distributions for the $\sigma_8$
marginalization intervals $[0.70,0.75]$ (first row), $[0.70,0.80]$
(second row), $[0.70,0.85]$ (third row), $[0.70,0.90]$ (fourth
row). These computations thus illustrate the effects of increasing
systematic errors without identifying the exact sources of the
systematics. In all cases the likelihood distributions have maxima at
$w_{\rm x}$ values between $-1.10$ and $-0.75$, and $\Omega_{\rm m}$ values
between 0.28 and 0.32.

The differences seen in Figs.\,\ref{FIG_NZSN1} and \ref{FIG_NZSN2} are
attributed to the different methods used to correct the SN peak
magnitudes, SN sample-to-sample variations, and different random plus
systematic errors in the reduction of the cluster data. In all cases
the centroids clearly fall between the NEC and SEC lines. Formal
averages over the mean values and their $1\sigma$ errors (without
cosmic variance) obtained with the largest marginalization range of
$\sigma_8=[0.70,0.95]$ give the final (most conservative) results,
\begin{equation}\label{FINAL}
\Omega_{\rm m}\,=\,0.29^{+0.08}_{-0.12}\,,\quad\quad w_{\rm x}\,=\,-0.95^{+0.30}_{-0.35}\,.
\end{equation}
Due to the statistical dependencies of the individual SN samples,
(\ref{FINAL}) gives the mean errors obtained with the individual
cluster-SN likelihood combinations, and not the errors of the averaged
$\Omega_{\rm m}$ and $w_{\rm x}$ values.

\section{Summary and conclusions}\label{DISCUSS}

\begin{table}
\renewcommand{\arraystretch}{1.5} {\bf Table\,1.} Constraints on
$\Omega_{\rm m}$ and $w_{\rm x}$, and their $1\sigma$ errors obtained with SNe
plus X-ray clusters of galaxies after marginalization over
$\sigma_8=[0.70,0.95]$ assuming a flat geometry. SNe: supernova
sample, LCC: light curve correction, MCLS: multi-color light curve
shape correction, $\Delta m_{15}$ correction, $s$: stretch factor
correction.\\
\vspace{-0.1cm}
\begin{center}
\begin{tabular}{lccc}
SNe    & LCC             & $\Omega_{\rm m}$              & $w_{\rm x}$ \\
\hline
Riess  & MCLS            & $0.28^{+0.09}_{-0.11}$ & $-0.75^{+0.25}_{-0.35}$\\
Riess  &$\Delta m_{15}$  & $0.30^{+0.08}_{-0.12}$ & $-1.05^{+0.35}_{-0.35}$\\
Perlmutter & $s$         & $0.30^{+0.08}_{-0.12}$ & $-1.05^{+0.30}_{-0.35}$\\
\hline
Average &                & $0.29^{+0.08}_{-0.12}$ & $-0.95^{+0.30}_{-0.35}$\\
\hline
\hline
\end{tabular}
\end{center}
\end{table}

The null energy condition (NEC) and the strong energy condition (SEC)
of general relativity are tested and give observational constraints on
cosmic phenomena like quintessence, super-quintessence, and Einstein's
cosmological constant. In order to test NEC and SEC on cosmic scales
we assume a flat geometry and that baryonic matter, non-baryonic
matter, relativistic matter and $\Lambda$-like matter can be regarded
as independent perfect fluids. The resulting inequalities
(Eqs.\,\ref{NEC} and \ref{SEC}) reproduce the frequently used
threshold $w_{\rm x}=-1$ only in the unrealistic limit of a zero density of
ordinary matter. Note that the usually adopted threshold is based on
the unknown component $\Omega_{\rm x}$ only, which is not enough to cover
the large diversity of phenomena caused by a mixture of different
cosmic fluids expected to fill the present Universe
(Fig.\,\ref{FIG_SCATCH}).

In this sense simple cosmological tests of NEC and SEC are formulated
and applied to the presently largest homogeneously selected sample of
X-ray cluster of galaxies (REFLEX). The most important advantage of
X-ray clusters as summarized in Borgani \& Guzzo (2001) is that in
contrast to optically selected clusters, their selection function
(e.g. the sample volume, see B\"ohringer et al. 2002) and the relation
between cluster X-ray luminosity and total cluster mass (see Reiprich
\& B\"ohringer 2002) is well-known without the need of extensive
numerical simulations. Moreover, the relation between the presence of
a cluster in X-rays and a peak in the underlying cosmic mass
distribution (cluster biasing) can be derived from first principles
(e.g. Kaiser 1994, Matarrese et al. 1997, Moscardini et
al. 2000). Therefore, the abundances of the nearby clusters are quite
sensitive to $\Omega_{\rm m}$ and almost independent of $w_{\rm x}$, which is
optimal to break the degeneracy between $\Omega_{\rm m}$ and $w_{\rm x}$ shown by
the SN data.

Recently, detailed studies showed that for the application of X-ray
clusters as cosmological probes several systematic errors could quite
strongly affect the final results (H. B\"ohringer et al., in
preparation, Pierpaoli et al. 2002, Reiprich \& B\"ohringer 2002,
Schuecker et al. 2003). This is reflected in a large scatter of
$\sigma_8$ values published recently by different groups using
different samples and methods and by comparing X-ray cluster results
with results obtained with, e.g., weak lensing, optical clusters,
X-ray cluster temperature and luminosity functions, power spectrum,
Sunyaev-Zel'dovich effect power spectrum, galaxy clustering etc. --
although some kind of convergence to specific $\Omega_{\rm m}$ and
$\sigma_8$ values emerges. The present investigation takes this
imprecise knowledge into account by a marginalization over a quite
large range of $\sigma_8$ values $[0.70,0.95]$ which includes about
75\% of the values obtained within the past two years. In comparison
to this range, additional marginalization over our imprecise knowledge
of cosmological parameters in the ranges $0.64\le h\le 0.80$,
$0.018\le \Omega_{\rm b}h^2\le 0.026$, and $0.8\le n_{\rm S}\le 1.2$ was
analysed and shown to be of secondary importance and is thus neglected
in the final results presented here.

It is quite important to note that the final $\Omega_{\rm m}$ and $w_{\rm x}$
values obtained in the present investigation are {\it almost
independent} on the assumed marginalization ranges. Different choices
in $\sigma_8$ yield changes in the final results always smaller than
about 7\% in both $\Omega_{\rm m}$ and $w_{\rm x}$. An increased marginalization
range soley increases the final error bars and leaves the centroid
values almost unchanged. We attribute the robustness of the test to
the complementarity of SN-Ia and X-ray cluster data. Future
investigations of $w_{\rm x}$ will clearly benefit from this
complementarity.

The present analysis neglects the $w_{\rm x}$-dependency of both the shape,
and the amplitude growth of the power spectrum of the matter density
fluctuations. This is justified by the limited spatial scale and
redshift ranges covered by the REFLEX cluster sample. Future cluster
samples will hardly reach $5\,h^{-1}\,{\rm Gpc}$ scales or so where
linear theory $w_{\rm x}$-dependent effects on the shape of the power
spectrum are formally expected. However, $w_{\rm x}$ can change the
amplitude of the power spectrum by a factor of about two between
redshift zero and $z=1$ which could be measured if the cluster X-ray
luminosity/mass conversion and the effective biasing parameter of the
sample could be computed with high enough accuracy. Deep and wide
X-ray cluster samples could thus use in addition to the mean cluster
abundance as the traditional cluster criterion (as described in
Haiman, Mohr \& Holder 2001) another quite strong $w_{\rm x}$-dependent
criterion related to the fluctuations of the cluster counts around
their mean abundance.

In the present investigation the joint likelihood distributions of the
combination of X-ray cluster and SN data are computed as a function of
the present-day $\Omega_{\rm m}$ and $w_{\rm x}$ values. NEC and SEC are thus
effectively tested at redshift zero.

The combined data fall with about $1.5-4\sigma$ statistical
significance (depending on the SN sample and light curve correction)
below the SEC threshold. The SEC can thus tendatively be regarded as
broken on cosmic scales at $z=0$. A similar breaking of SEC was found
by Visser (1997) under more general assumptions for the time between
the epoch of galaxy formation and the present where he compares the
relation between the age of the Universe and the age of the oldest
observed stars. However, our theoretical expectations suggest that the
broken SEC state should not hold for redshifts $z>(0.28$-$1.20)$
(Eq.\,\ref{EC} and the $1\sigma$ error corridor given in
Eq.\,(\ref{FINAL})), in contrast to the larger $z$ range implied by
the analysis of Visser.

The data fall above the NEC threshold with about $1.5-3\sigma$
statistical confidence, again depending on the SN sample and
light-curve correction used. NEC can thus tendatively be regarded as
fulfilled. 

In every case tested sofar, the combined X-ray cluster and SN data
obviously populate the sector between NEC validation and SEC violation
and thus provide further observational evidence for an accelerated
cosmic expansion at $z=0$.

The observational constraints obtained from the combination of X-ray
cluster and SN data on $\Omega_{\rm m}$ are in good agreement with
recent cluster data (e.g. Borgani et al. 2001, Allen et al. 2002,
Schuecker et al. 2002, 2003, Pierpaoli et al. 2002), constraints from
CMB data (see references given in Sect.\,\ref{INTRO}, especially the
WMAP result $\Omega_m=0.29\pm 0.07$, 68\% confidence, WMAP data only,
Spergel et al. 2003), and galaxy data (Szalay et al. 2001, Lahav et
al. 2002). 

Our results on $w_{\rm x}$ are consistent with the constraints
obtained from type-Ia supernovae (Garnavich et al. 1998, Perlmutter et
al. 1999), with recent CMB data (e.g., Baccigalupi et al. 2002, Bond
et al. 2002) and with the baryonic fraction in galaxy clusters (Ettori
et al. 2003).  The results obtained by the combined analysis of CMB
and SN data of Hannestad \& M\"ortsell (2002) and Melchiorri et
al. (2002) yield the respective 95\% confidence constraints
$-2.68<w_{\rm x}<-0.78$ and $-1.62<w_{\rm x}<-0.74$ (see also Caldwell
2002), quite consistent but slightly larger than the results obtained
with the SN and X-ray cluster data given here. Finally, the constraint
$w_{\rm x}\le -0.78$ is obtained with 95\% confidence from the
combination of WMAP, SN, 2dFGRS and Ly$\alpha$ data (Spergel et
al. 2003).

Type-Ia SN and X-ray cluster data thus support a picture of a universe
which is presently in a state of accelerated expansion, where NEC is
most probably not violated (no super-acceleration with a possible
catastrophic ending, McInnes 2002), and in which a cosmological
constant or something like it provides the dark energy.

However, one still has to be cautious with conclusions about NEC and
SEC because they are only tested under restricted conditions.
Furthermore, we are aware of the necessity to study in much more
detail our assumptions that both galaxy clusters and SNe\,Ia do not
evolve over the respective redshift ranges covered by the given
observations. Moreover, important relations like the total cluster
mass/X-ray luminosity relation and the extinction law relevant for
nearby and distant SNe have to be known with much higher precision
because the deviations from $w_{\rm x}=-1$ at $z=0$ for interesting
dark energy scenarios might be smaller than the error bars of the
present results. The present investigation tries to take into account
possible systematic errors of the treatment of the X-ray clusters by
using a quite large $\sigma_8$ marginalization interval. Future
measurements based on improved relations can work with smaller
intervals expected to provide quite precise cosmological constraints
on both $\Omega_{\rm m}$ and $w_{\rm x}$.

\begin{acknowledgements}                                                        
We thank Stefan Gottl\"ober and Stefano Ettori for useful
discussions. PS acknowledges financial support under grant
No.\,50\,OR\,0108. RRC thanks the Santa Barbara KITP for
hospitality. This work was supported at the KITP by NSF PHY99-07949,
and at Dartmouth by NSF PHY-0099543.
\end{acknowledgements}

\end{document}